\documentclass{aa}

\usepackage[varg]{txfonts}
\usepackage{natbib}
\bibpunct{(}{)}{;}{a}{}{,}

\usepackage{hyperref}
\usepackage{graphicx}
\usepackage{booktabs}
\usepackage{multirow}
\usepackage{xcolor}
\usepackage[separate-uncertainty=true, multi-part-units=single]{siunitx}
\usepackage{threeparttable}

\newcommand{\fbin}{\ensuremath{f_\text{bin}}}

\newcommand{\Kepler}{\textit{Kepler}}

\DeclareSIUnit \Msun {\ensuremath{\text{M}_\odot}}
\DeclareSIUnit \Rsun {\ensuremath{\text{R}_\odot}}
\DeclareSIUnit \Lsun {\ensuremath{\text{L}_\odot}}
\DeclareSIUnit \Tsun { \ensuremath{ \text{T}_{ \text{eff}, \odot } } }
\DeclareSIUnit \mag {mag}
\DeclareSIUnit \pc {pc}
\DeclareSIUnit \yr {year}

\makeatletter
\renewcommand*\aa@pageof{, page \thepage{} of \pageref*{LastPage}}
\makeatother

\begin{document}

\title{DUETS: Setting expectations for asteroseismic binaries and binary products with synthetic populations}

\author{
    A. Mazzi\inst{1,2} \fnmsep \thanks{alessandro.mazzi@hotmail.com}
    \and J. S. Thomsen\inst{1,3,4}
    \and A. Miglio\inst{1,3,5}
    \and K. Brogaard\inst{4,1}
    \and L. Girardi\inst{2}
    \and D. Bossini\inst{6,2}
    \and M. Matteuzzi\inst{1}
    \and W. E. van Rossem\inst{1}
}

\institute{
    Department of Physics and Astronomy "Augusto Righi", University of Bologna, via Gobetti 93/2, 40129, Bologna, Italy
    \and INAF-Osservatorio Astronomico di Padova, Vicolo dell'Osservatorio 5, I-35122 Padova, Italy
    \and INAF-Astrophysics and Space Science Observatory of Bologna, via Gobetti 93/3, 40129, Bologna, Italy
    \and Stellar Astrophysics Centre, Department of Physics and Astronomy, Aarhus University, Ny Munkegade 120, 8000, Aarhus C, Denmark
    \and School of Physics and Astronomy, University of Birmingham, Edgbaston, B15 2TT, UK
    \and Dipartimento di Fisica e Astronomia Galileo Galilei, Università di Padova, Vicolo dell'Osservatorio 3, 35122, Padova, Italy
}

\titlerunning{DUETS: Setting expectations for asteroseismic binaries and binary products with synthetic populations}

\authorrunning{Mazzi et al.}

\date{}

\abstract
{Binary stars and products of binary evolution contribute significantly to shaping stellar populations, yet are often neglected due to being difficult to identify. With asteroseismology providing precise stellar parameters, we have an opportunity to better characterize binaries and their products, and refine our understanding of their role in Galactic populations.}
{We estimate occurrence rates, mass distributions, and evolutionary states of asteroseismic binaries, exhibiting solar-like oscillations from both components, and of products of binary interactions with detectable solar-like oscillations. Additionally, we explore the effects of mass accretion or loss on ``apparent age''-metallicity relations.}
{We simulate \SI{121}{\square\deg} of \Kepler{}'s field of view using the TRILEGAL population synthesis code, adopting the Eggleton and Moe \& Di Stefano distributions of initial binary parameters, and generate an additional simulation with non-interacting binaries for comparison.}
{We find that asteroseismic binaries require an initial mass ratio close to one, and even small mass transfer events can prevent the detection of oscillations from both components. The non-interacting case yields the highest fraction of asteroseismic binaries for red giant stars with detectable oscillations (0.46\%), while Eggleton yields the lowest (0.06\%). Asteroseismic binaries composed of two red clump stars are not expected at separations smaller than \SI{500}{\Rsun} due to the interplay of stellar evolution and binary interactions. The simulation with the Moe \& Di Stefano distribution suggests that $\sim 1\%$ of \Kepler{}'s field red giants with detectable oscillations have undergone significant mass accretion or loss, appearing rejuvenated or prematurely aged and potentially affecting Galactic age-metallicity relations, although the occurrence and properties of these populations strongly depend on the assumed initial binary parameters.}
{Comparing predicted and observed asteroseismic binaries, as well as over- and under-massive stars, offers a way to constrain key binary evolution assumptions, such as the initial binary fraction and period distribution, and to reduce uncertainties in mass-transfer modeling.}


\maketitle
\nolinenumbers

\section{Introduction}
Binary stars are a frequent occurrence in the Milky Way \citep{eggleton06, moedistefano2017}, and almost certainly the same is true in all external galaxies.
Stars in binary systems are generally assumed to have a common origin and therefore have the same initial chemical composition.
They orbit around their common center of mass and can interact in different ways, thereby altering their evolutionary pathway \citep{demarco17_binaryevolution}.
For example, dynamical interactions can influence their orbit and the rotational velocity of the components \citep[see for instance][]{pricewhelan18_circularization}.
If they are close enough, mass can be transferred from the less dense component to the other one, potentially stripping the former of its external layers and revealing the stellar interior \citep{demink14, handberg17, brogaard21_ngc6791, li22, ramachandran23, geier24_hotsubdwarfs}.
During interaction, the binary stellar components might also undergo one or more common envelope events and potentially merge, creating a single, more massive \cite{paczynski1976, ivanova13, ropke23} or less massive \citep{matteuzzi24} star.
All these mechanisms eventually lead to stars with properties that cannot be explained using the standard theory of stellar evolution.

Finding and characterizing binary systems is crucial to put strong constraints on both single- and binary-star evolution.
In the literature a multitude of methods have been used to find binaries across different ranges of stellar and orbital parameters, as discussed in \cite{moedistefano2017}.

However, the identification and characterization of binaries and products of binary evolution can be a difficult task as different methods are generally sensitive only to a small subset of the orbital configurations \cite{moedistefano2017}.
In this context, asteroseismology is a powerful tool as it offers the opportunity to determine with high precision fundamental parameters of stars, including mass, radius, and evolutionary stage, across various types of binary systems, and to uncover objects resulting from non-standard stellar evolution.
In fact, studies of oscillations in stars as they appear on the surface, i.e. through Fourier analysis of photometric or spectroscopic time series, has enabled investigations of the stellar interiors for a variety of pulsator types.
The most well studied are the ``solar-like'' oscillators, where the convection in the outer layers stochastically drives and dampens the resonant modes of the star.
This class of oscillators comprises low-mass main sequence (MS) stars as well as red giant (RG) stars.
The information that we can learn through asteroseismology, on top of the mass and radius of the star, comprises e.g. its evolutionary state, the mass of the core, the He abundance, and the internal rotation profile \citep[see for instance][]{garcia19}.

The \Kepler{} mission has produced long, high-quality photometric time series for numerous targets in a region of the sky of about $100$ deg$^2$ (hereafter mentioned as \Kepler{}'s field).
The 30-min cadence\footnote{The Nyquist frequency for this cadence is approximately \SI{280}{\micro\Hz}.} of \Kepler{} photometry and \SI{4.5}{years} of continuous observing, leading to a frequency resolution of approximately \SI{0.008}{\micro\Hz}, represented the perfect opportunity to study core-He burning stars and low-luminosity red giant branch (RGB) stars, which typically have oscillation frequencies between \num{20} and \SI{200}{\micro\Hz} \citep[see e.g. figure~6 of][]{miglio09}.

With photometric time-series, binaries are typically detected through either eclipses, which require that the systems are observed nearly edge-on, or through tidal deformation signals.
These methods, however, have an observational bias toward shorter-period binaries.
Eclipses become rarer with larger semi-major axis as the allowed inclination range narrows, and require longer photometric time-baselines to assure detection.
Tidal signals similarly decrease sharply in amplitude with increasing semi-major axis.
Radial velocities can detect systems with longer periods (i.e., $\gtrsim \SI{1e4}{\day}$), but require extended time baselines, and the signal strength depends on the orbital period and is highly sensitive to the system’s inclination.

Asteroseismology offers the possibility of overcoming some of the limitations of the previously mentioned methods.
In case the system hosts two stars that show solar-like oscillations, the frequency power spectrum will contain two oscillation envelopes, at frequencies determined by the properties of the two components, and this is largely independent from the orbital period of the binary system.
Binary systems in which oscillations for both stars are detected are called ``asteroseismic binaries``.

\cite{miglio14} investigated the binary content of \Kepler{}'s field through simulations performed with two different population synthesis codes, TRILEGAL \citep{girardi05, pieres20} and BiSEPS \citep{willems02_biseps}.
The authors focused the discussion on the results of the former, which at that time did not include prescriptions for simulating interactions during the evolution of binary stars, and instead the components evolved following traditional single-star evolution.

To date, a systematic search for asteroseismic binaries or products of binary evolution among solar-like oscillations has not yet been performed, with only few cases reported in the literature \citep[e.g.][]{appourchaux15, white17, beck18, li18, matteuzzi23, matteuzzi24}.
We are currently conducting such a search, the results of which will be presented in a forthcoming companion paper.

In this work, instead, we aim to provide predictions of occurrence rates, mass distributions, evolutionary states, and chemical compositions of asteroseismic binaries and products of coalescence or mass exchange that exhibit detectable oscillations.
Crucially, we intend to investigate the impact of different assumptions about the underlying binary population, including the effects of binary interactions.
Furthermore, we aim to estimate how over- or under-massive giants might influence the 'apparent' age-metallicity relations, extending beyond the case of thick disk stars (as discussed, e.g., in \citealt{izzard18}).

The paper is structured as follows.
In Section~\ref{sec:simulation} we describe the population synthesis model we adopt and the different prescriptions we use to generate the initial parameters of the binary stars, as well as the setup related to the region of the sky we simulate.
Section~\ref{sec:results} presents the results for one simulation and describes the binary stars and products of binary interactions that have been produced.
In Section~\ref{sec:comparison} we compare these results with those obtained using different binary initial parameter prescriptions and in \ref{sec:finding} we discuss briefly how asteroseismology can identify binary stars and products of their evolution.
Finally, we give our conclusions in Section~\ref{sec:conclusion}.

\section{Simulation}
\label{sec:simulation}
Several numerical codes are available in the literature to generate mock catalogs of stellar populations including binaries, for example binary\_c \citep{izzard04_binaryc, izzard06, izzard09, izzard18, izzard23}, BiSEPS \cite{willems02_biseps}, and BSE \citep{hurley02_bse}.
In this work we use the TRILEGAL population synthesis code and specifically its BinaPSE module.
We briefly describe these software programs below, and direct the interested reader to their primary publications for more detailed information.

\subsection{TRILEGAL}\label{sec:trilegal}
The TRIdimensional modEl of the GALaxy \citep[TRILEGAL][]{girardi05, vanhollebeke09, girardi12, pieres20} is a population synthesis code that can generate synthetic populations of stars in the Milky Way, in a stellar cluster, or even in external galaxies\footnote{Online interfaces for the code are available at \url{http://stev.oapd.inaf.it/cgi-bin/trilegal} and \url{http://stev.oapd.inaf.it/cgi-bin/cmd}.}.

A TRILEGAL simulation of a given field of view, in our case an approximation of the \Kepler{} field (see Section~\ref{sec:sim-setup}), starts by first drawing a large sample of stars from pre-defined distributions of mass, age and metallicity.
The number of stars in the simulation is determined by the area of the field of view and the density integrated along the line of sight of the Milky Way's components that are implemented in TRILEGAL \citep[bulge, thin and thick disk, halo, see also][]{vanhollebeke09, pieres20}.
Each component's density distribution is also employed to determine the distance along the lines of sight of each star.
Initial masses are drawn from the initial mass function (IMF), while ages and metallicities come from the star formation rate (SFR) and age metallicity relation (AMR) of each galactic component, in proportion to their contribution to the field's stellar density.
Then, a grid of stellar evolutionary tracks computed with the PARSEC stellar evolution code \citep{bressan12_parsec,bressan13} is employed to determine the evolution of each star and derive additional stellar properties, such as effective temperature, luminosity, surface gravity and evolutionary state.
Extinction is computed with TRILEGAL's dust model\footnote{TRILEGAL assumes that the dust layer in our galaxy is aligned with the galactic plane and has a density that is exponentially decreasing with distance from the Galactic plane, see also \cite{girardi05}.}, normalized at infinity using the extinction provided by the Planck 2D extinction map \citep{planckmap14} at the location of the field, and is included in the computation of the bolometric corrections \citep{chen19_ybc} to derive apparent magnitudes in the \Kepler{} photometric passband.
The output of the simulation contains the initial and final properties of the stars.
Further details on the simulation setup can be found in Sect.~\ref{sec:sim-setup}.

\subsection{BinaPSE}\label{sec:binapse}
    Above we have outlined the procedure to generate a population of single stars.
    Simulating binary systems, while accounting for potential binary interaction, requires further steps which have been implemented in the TRILEGAL code with the BinaPSE module \citep{daltio21}.
    The generation of binary systems starts, like in the case of single stars, by drawing the initial mass of a star from an IMF, which is considered to be the primary star of the system, i.e the more massive component.
    Then, the mass ratio, and therefore the initial mass of the secondary star, the less massive component, as well as the initial elliptical orbital parameters are drawn assuming a distribution of initial binary properties.
    In this work we explore the impact of the \cite{eggleton06} and \cite{moedistefano2017} distributions of initial parameters, which we present in more detail in Section~\ref{sec:initial-params-binary}.

The components of the binary systems are evolved following single star evolution except during stages where binary interactions are determined to occur.
Using a time-step small enough to avoid abrupt changes of stellar and orbital parameters, the orbit is solved with an analytic orbit calculator and the potential effects of binary interaction are determined using analytic prescriptions for mass loss, mass transfer via Roche lobe overflow or winds, common-envelope evolution, or changes in the orbital parameters due to tidal effects.
If a change in the mass of a component of the binary system occurs due to binary interaction, a new stellar track will be selected from the grid, matching its current mass.
The outcomes of common-envelope phases, mergers or stripping follow the well-established set of prescriptions implemented in the Binary Stellar Evolution (BSE) code \citep{hurley02_bse}.
However, BinaPSE differs from BSE by using the grids of stellar tracks and interpolation techniques of TRILEGAL in place of the analytic stellar evolution prescriptions of BSE.
The set of evolutionary tracks has also been expanded with BinaPSE to be able to simulate more exotic binary evolution results, as described in \citep{daltio21}.

\subsection{Initial parameters for the binary stars}\label{sec:initial-params-binary}
To simulate a stellar population with BinaPSE it is necessary to choose distributions for the initial parameters of the binary systems, such as mass ratio and the initial orbital parameters.
In the current version of the software, two distributions are implemented in BinaPSE: the Monte Carlo model outlined in \citet[hereafter E06]{eggleton06} and the model described in \cite{moedistefano2017} (MDS17).
We briefly describe both in the following.

\paragraph{E06.}
This prescription assumes that the mass $M_1$ of the primary star at the beginning of the simulation, i.e. the initially most massive star in the binary, has been generated according to a given initial mass function (IMF).
Then, the mass of the companion, $M_2 \leq M_1$, and all other parameters of the binary system can be determined according to the distributions listed in the following.
The initial orbital period of the binary is computed from
\begin{equation}\label{eq:egg_P}
    P = \frac{5 \times 10^4}{M_1^2} \left( \frac{X_P}{1-X_P}\right)^\alpha,
\end{equation}
with
\begin{equation}
    \alpha = \frac{3.5 + 1.3 \alpha^\prime}{1+\alpha^\prime}
\end{equation}
and
\begin{equation}
    \alpha^\prime = 0.1 M_1^{1.5},
\end{equation}
and $X_P$ a random number uniformly distributed in $[0,1]$.
Similarly, the mass ratio of the binary $q = M_2/M_1$, with $0 < q \leq 1$, is obtained from
\begin{equation}
    q = 1-X_q^\beta
\end{equation}
with
\begin{equation}
    \beta = \frac{2.5 + 0.7 \beta^\prime}{1 + \beta^\prime}
\end{equation}
and
\begin{equation}
    \beta^\prime = 0.1 P^{0.5} \left( M_1 + 0.5 \right)
\end{equation}
where again $X_q$ is a random number uniformly distributed in $[0,1]$.
Finally, the eccentricity of the binary is determined from
\begin{equation}
    e = X_e
\end{equation}
where $X_e$ is a random number uniformly distributed in $[0,1]$.

\paragraph{MDS17}
In contrast to the E06 prescription, where the parameters depend just on the mass of the primary star, the MDS17 parameter distributions account for correlations between the parameters, providing different relations for specific regions of the parameter space.
We avoid a full description of the model here and refer to the MDS17 paper for all the details.
However, we note that the orbital periods in the MDS17 model are limited to the interval between \SI{\sim1.6}{\day} and \SI{e8}{\day}, while in the E06 model they are not constrained.
As will be discussed in Section~\ref{sec:results-over}, this leads to a significant difference in the number of systems that interact and merge.

\paragraph{Non-interacting binaries} An additional model that can be used to generate the initial parameters of binary stars is the non-interacting binary star model originally implemented in TRILEGAL.
For every single star generated by TRILEGAL, with mass taken from a given IMF, this model assumes that there is a probability equal to the user-defined binary fraction value that the star has a companion.
The mass ratio of each system is randomly generated from a uniform distribution between a user-defined minimum value $q_\text{min} = 0.7$ and the maximum $q_\text{max} = 1$, while the orbital parameters are not simulated. The two components of the binaries are then evolved using only single stellar evolution, as if isolated, and therefore no interaction, either dynamical or via mass transfer, can occur.

\subsection{Simulation setup and runs}
\label{sec:sim-setup}
We have performed a simulation with each distribution of initial parameters presented above for a region with an area of $\sim\SI{121}{\square\deg}$ centered on $(l,b)=(76.3, 13.5)$, without using a detailed description of the field of view of \Kepler{}.
The area of the simulation is also divided into smaller regions according to the HEALPix\footnote{\url{https://healpix.sourceforge.io/}} scheme \citep{gorski05_healpix} with an NSIDE of 64, through the \texttt{healpy}\footnote{\url{https://github.com/healpy/healpy}} Python package \citep{zonca19_healpy}, to account for the variation across the sky of extinction and stellar density, the latter being quite significant in such a low-latitude field.
The extinction values themselves are taken from the 2D extinction map produced by the Planck mission \citep{planckmap14} after degrading its initial resolution (NSIDE=2048) to match the one used in the simulation.
We simulate only the field stars, which are part of the thin and thick disks and the halo, and do not simulate any of the clusters in \Kepler{}'s field of view.
Regarding the binary content, for the simulations adopting the MDS17 and E06 prescriptions\footnote{We will refer to the simulations produced with these prescriptions as MDS17 simulation and E06 simulation, respectively.} we produce populations that have a fraction $\fbin=0.3$ of their initial mass that goes into the generation of binary systems, and we do not attempt to constrain this value in the present work.
For the simulation of the non-interacting binaries we also adopt $\fbin=0.3$ and, although in this case it is interpreted as the probability of a star to have a companion, we check that all three simulations have an initial number fraction of binaries that is roughly 40\%.
We have also run a simulation with the MDS17 prescription and $\fbin=0.6$, and observed that the number of binaries in the phases listed in Table~\ref{tab:counts_binaries} roughly doubles.
All other parameters in the simulation are left to their default values as indicated in the main TRILEGAL papers \citep{girardi05, vanhollebeke09, girardi12, pieres20}.

Regarding the photometric depth, we chose to adopt the same limit as in the \Kepler{} Input Catalog \citep[KIC, ][]{koch10}, that is, \SI{\sim 16}{mag} in the \Kepler{} band.
For binary stars, the limit is applied taking into account their total magnitude.

Finally, contrary to BinaPSE, in the current version of TRILEGAL the definition of primary and secondary components refers to the most and least massive component of the system at the end of the simulation rather than at the beginning.
In the following sections, we adopt the TRILEGAL definition.

\section{Results}
\label{sec:results}

\begin{table*}
    \caption{Counts of stars in different evolutionary stages for the simulation adopting the \citet{moedistefano2017} prescription.}
    \label{tab:counts_phases}
    \centering
    \begin{tabular}{lrrrrrr}
        \toprule
        phase\tablefootmark{a}
        & N
        & N$_\text{sin}$
        & N$_\text{lone pri}$
        & N$_\text{pri}$ & N$_\text{sec}$
        & (N$_\text{pri}$ + N$_\text{sec}$) / N \\
        \midrule
        MS & 263720 & 113686 & 6 & 68525 & 81503 & 0.56889 \\
        HG & 27998 & 16648 & 4 & 10236 & 1110 & 0.40524 \\
        RGB & 35502 & 22533 & 102 & 12461 & 406 & 0.36243 \\
        CHeB & 20495 & 12629 & 730 & 6648 & 488 & 0.34818 \\
        EAGB & 2527 & 1744 & 54 & 675 & 54 & 0.28848 \\
        TP-AGB & 143 & 121 & 4 & 18 & 0 & 0.12587 \\
        Post-AGB & 42 & 5 & 2 & 29 & 6 & 0.83333 \\
        CO-WD & 7894 & 0 & 0 & 183 & 7711 & 1.00000 \\
        He-WD & 8041 & 0 & 0 & 127 & 7914 & 1.00000 \\
        \bottomrule
    \end{tabular}
    \tablefoot{
        ``N'' indicates the total number of stars, ``N$_\text{sin}$'' the number of single stars, ``N$_\text{lone pri}$'' the number of single stars remainders of binaries (due to a merger of the ejection of the companion), and ``N$_\text{pri}$'' and ``N$_\text{sec}$'' refer to primaries and secondaries stars in binaries, respectively.\\
        \tablefoottext{a}{Main sequence (MS), Hertzsprung gap (HG), red giant branch (RGB), core-Helium burning (CHeB), early-asymptotic giant branch (EAGB), thermal-pulsations AGB (TP-AGB), post-AGB, Carbon-Oxygen white dwarf (CO-WD) and Helium white dwarf (He-WD).}
    }
\end{table*}

\begin{table*}
    \caption{Similar to Table~\ref{tab:counts_phases}, but limited to stars with detectable oscillations.}
    \label{tab:counts_phases_det}
    \centering
    \begin{tabular}{lrrrrrrrrr}
        \toprule
        phase & N$^\text{det}$ & N$^\text{det}_\text{sin}$ & N$^\text{det}_\text{lone pri}$ & N$^\text{det}_\text{pri}$ & N$^\text{det}_\text{sec}$ & (N$^\text{det}_\text{pri}$ + N$^\text{det}_\text{sec}$) / N$^\text{det}$ & N$^\text{seismo}_\text{pri}$ & N$^\text{seismo}_\text{sec}$ & (N$^\text{seismo}_\text{pri}$ + N$^\text{seismo}_\text{sec}$) / N$^\text{det}$ \\
        \midrule
        MS & 7528 & 5381 & 0 & 2099 & 48 & 0.28520 & 30 & 33 & 0.00837 \\
        HG & 3262 & 2205 & 1 & 1038 & 18 & 0.32373 & 7 & 5 & 0.00368 \\
        RGB & 21227 & 15046 & 91 & 5952 & 138 & 0.28690 & 49 & 24 & 0.00344 \\
        CHeB & 17906 & 12307 & 703 & 4653 & 243 & 0.27343 & 58 & 81 & 0.00776 \\
        EAGB & 2354 & 1743 & 53 & 511 & 47 & 0.23704 & 7 & 8 & 0.00637 \\
        TP-AGB & 139 & 119 & 4 & 16 & 0 & 0.11511 & 0 & 0 & 0.00000 \\
        Post-AGB & 23 & 1 & 2 & 15 & 5 & 0.86957 & 0 & 0 & 0.00000 \\
        CO-WD & 0 & 0 & 0 & 0 & 0 & 0.00000 & 0 & 0 & 0.00000 \\
        He-WD & 0 & 0 & 0 & 0 & 0 & 0.00000 & 0 & 0 & 0.00000 \\
        \bottomrule
    \end{tabular}
    \tablefoot{
        Total number of stars (``N$^\text{det}$''), single stars (``N$_\text{sin}^\text{det}$''), lone primaries (``N$_\text{lone pri}^\text{det}$''), primaries (``N$_\text{pri}^\text{det}$'') and secondaries (``N$_\text{sec}^\text{det}$'') with detectable oscillations. ``N$_\text{pri}^\text{seismo}$'' and ``N$_\text{pri}^\text{seismo}$'' are the number of stars in each evolutionary phase that are in asteroseismic binaries as primaries or secondaries, respectively.
    }
\end{table*}

In this section, we present key properties of the simulation using the MDS17 initial distributions of binary parameters.
We briefly discuss the general contents of the simulations and then focus on the binary stars, using the single stars as a baseline for the expected behavior of stars in isolation.
We compare these results with the E06 and non-interacting binary simulations in Section~\ref{sec:comparison}.

\subsection{Overall contents of the simulation}
In Table~\ref{tab:counts_phases} we give a summary of the stellar counts in different evolutionary phases for the MDS17 simulation, following the magnitude cut outlined in Section~\ref{sec:sim-setup}.
We count separately the true single stars (N$_\text{sin}$) and the single stars that started as binaries (N$_\text{lone pri}$), which we will refer to as ``lone primaries''\footnote{Tables and figures will adopt the abbreviated term ``lone pri'' to optimize space usage.}

As a result of the magnitude cut for binaries being applied to the total magnitude of the system, the counts for primary and secondary components, $\text{N}_\text{pri}$ and $\text{N}_\text{sec}$ respectively, include stars that are individually dimmer than \SI{16}{mag} in \Kepler{}'s band.

\subsection{Detection of solar-like oscillations}\label{sec:detection-sin}
The detection of solar-like oscillations in the power spectrum of a star depends on its intrinsic properties, as well as on its apparent magnitude, the duration of the observations and the characteristic noise of the instrument.
While TRILEGAL does not simulate stellar oscillations, \cite{chaplin11} presents a relatively simple procedure for determining the probability of detecting an asteroseismic signal in the power spectrum of a star from its fundamental stellar parameters, namely effective temperature, radius, mass and \Kepler{} apparent magnitude of a given star, and the length of the \Kepler{} data available to compute the power-spectrum.
We consider stars to be asteroseismically detectable if the probability of detecting oscillations using the complete 4-year dataset from \Kepler{} is at least 90\% \citep[e.g., see][]{Chaplin2011b}.

Table~\ref{tab:counts_phases_det} presents the counts of stars with detectable oscillations obtained from the simulation.
A small fraction of single MS and Hertzsprung gap (HG) stars have detectable oscillations, $4.7\%$ and $13.2\%$ respectively, while about $66.9\%$ of the single RGBs and $97.4\%$ of the single core-He burning\footnote{With this term we identify both stars that end up in the red clump region of the HRD after undergoing the He flash, that we refer to as ``red clump stars'', as well as those that ignite He in non-degenerate conditions, which we refer to as ``secondary clump stars''.} (CHeB) ones show an asteroseismic signal.
The difference in the fraction of detectable RGB and CHeBs is related to the large luminosity range covered by RGB stars.
In fact, oscillations in low-luminosity RGB stars have small amplitudes and are thus detectable only up to a relatively short distance.

Given the rather complex target selection function of the \Kepler{} mission \citep{farmer13}, which was not reproduced in the simulations, we do not aim to precisely reproduce the number of stars in the \Kepler{} catalog.
However, we note that the number of red giants with detectable oscillations in the simulation ($\sim 42000$) is comparable to the reported detections ($\sim30000$, Garcia et al., in preparation), indicating reasonable consistency.
On the other hand, the number of stars with detectable oscillations on the MS is not taking into account effects that can suppress oscillations, such as magnetic activity \citep[see e.g.][]{chaplin11b_activity}.

Importantly, in the following we focus specifically on detecting solar-like oscillations, therefore we do not make predictions for other types of pulsators.
Although our simulations could be adapted for such purposes, it is beyond the scope of the present work.

\subsection{Detectability of asteroseismic binaries}\label{sec:seis_bin}

\begin{table*}
    \caption{Binaries with detectable oscillations produced by the simulation with the \cite{moedistefano2017} distribution of binary initial parameters. All binaries are considered unresolved.}
    \label{tab:counts_binaries}
    \centering
    \begin{tabular}{llrrrrrrr}
        \toprule
        phase$_\text{late}$ & phase$_\text{early}$ & N & N$_\text{one}$ & N$_\text{seismo}$ & N$_\text{seismo}$ / N$_\text{late}^\text{det}$ & N$_\text{seismo}$  / N$_\text{early}^\text{det}$ & N$_\text{seismo}$ / $\sum$ N$_\text{late}^\text{det}$ & N$_\text{seismo}$/$\sum$N$_\text{seismo}$ \\
        \midrule
        MS & MS & 59228 & 2070 & 30 & 0.00399 & 0.00399 & 0.00057 & 0.19868 \\
        HG & MS & 7620 & 1023 & 2 & 0.00061 & 0.00027 & 0.00004 & 0.01325 \\
         & HG & 521 & 9 & 5 & 0.00153 & 0.00153 & 0.00010 & 0.03311 \\
        RGB & MS & 8738 & 5589 & 1 & 0.00005 & 0.00013 & 0.00002 & 0.00662 \\
         & RGB & 257 & 133 & 20 & 0.00094 & 0.00094 & 0.00038 & 0.13245 \\
        CHeB & RGB & 145 & 69 & 29 & 0.00162 & 0.00137 & 0.00055 & 0.19205 \\
         & CHeB & 65 & 2 & 53 & 0.00296 & 0.00296 & 0.00101 & 0.35099 \\
        EAGB & RGB & 12 & 8 & 3 & 0.00127 & 0.00014 & 0.00006 & 0.01987 \\
         & CHeB & 6 & 2 & 4 & 0.00170 & 0.00022 & 0.00008 & 0.02649 \\
         & EAGB & 4 & 0 & 4 & 0.00170 & 0.00170 & 0.00008 & 0.02649 \\
        \bottomrule
    \end{tabular}
    \tablefoot{
        Columns ``phase$_\text{late}$'' and ``phase$_\text{early}$'' indicate the evolutionary phase of the most (late) and least (early) evolved star in the system, respectively. ``N'' is the total number of systems with the given phases for the most and least evolved star at the end of the simulation. ``N$_\text{one}$'' indicates the number of systems where only one component has detectable oscillations and ``N$_\text{seismo}$'' represents the counts of asteroseismic binaries. ``N$_\text{seismo}$ / N$_\text{late}^\text{det}$'' and ``N$_\text{seismo}$  / N$_\text{early}^\text{det}$'' are the fraction of asteroseismic binaries with respect to all stars with detectable oscillations in the phase of the most and least evolved star, respectively, and ``N$_\text{seismo}$ / N$_\text{late}^\text{det}$'' is also plotted in Figure~\ref{fig:hist_detectable_bin}. Finally, ``N$_\text{seismo}$ / $\sum$N$^\text{det}$'' represents the fraction of asteroseismic binaries to the total number of stars with detectable oscillations, regardless of their evolutionary stage, and ``N$_\text{seismo}$/$\sum$N$_\text{seismo}$'' is the fraction of each kind of asteroseismic binary.
    }
\end{table*}

\begin{figure}[t]
    \centering
    \includegraphics[width=\columnwidth]{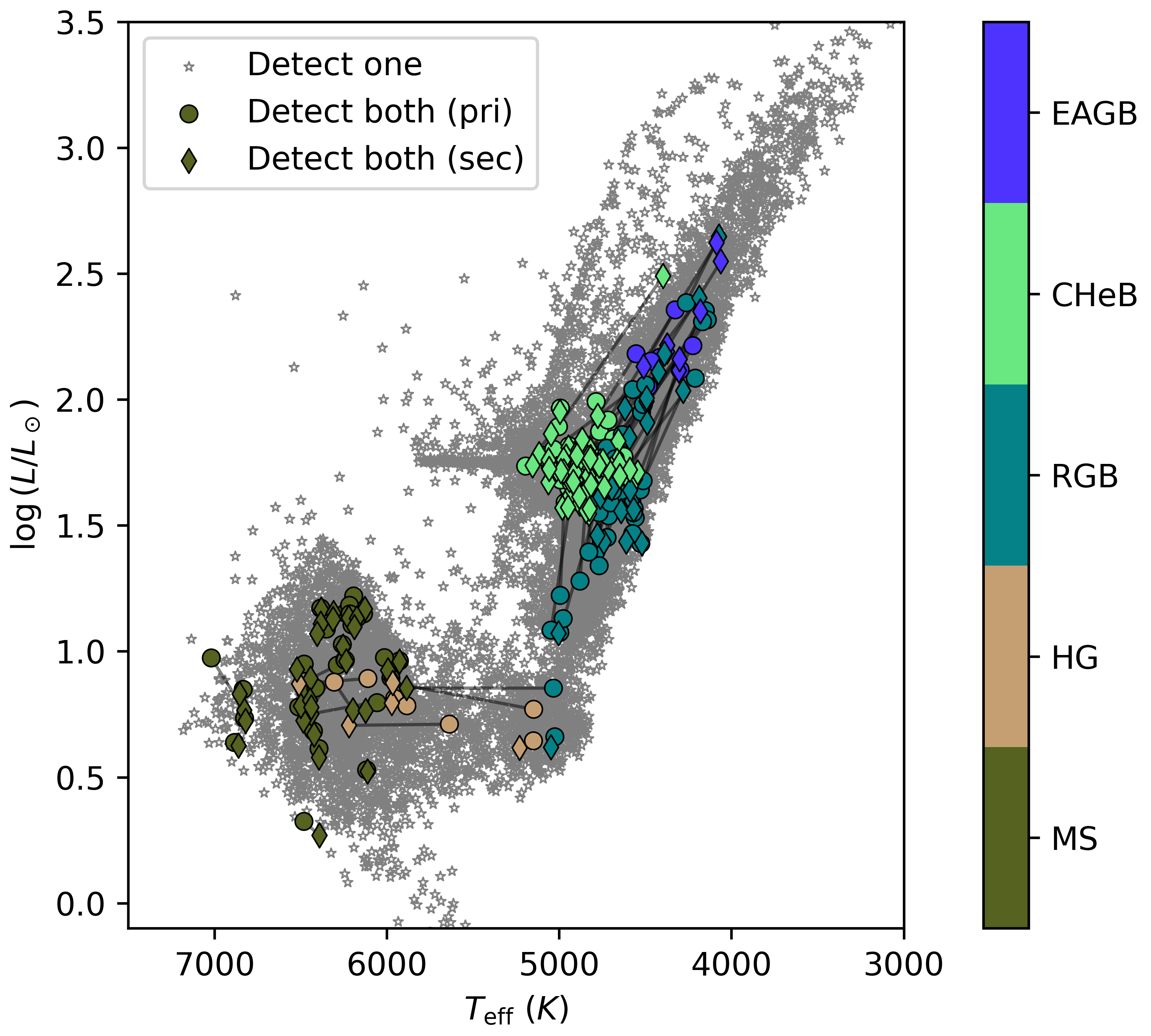}
    \caption{Hertzsprung-Russell diagram of the asteroseismic binaries produced by the simulation using the MDS17 prescription (Table~\ref{tab:counts_phases_det} and \ref{tab:counts_binaries}). The star symbols mark binary stars where only one component can be detected, and their luminosity corresponds to the total one of the binary star. The colored points represent the components of binary stars with detectable oscillations: the circle shows the location of the primary, while the thin diamond indicates the secondary. Thin black lines link together the components of the same binary.}
    \label{fig:hrd_detectable_bin}
\end{figure}

\begin{figure}[t]
    \centering
    \includegraphics[width=\columnwidth]{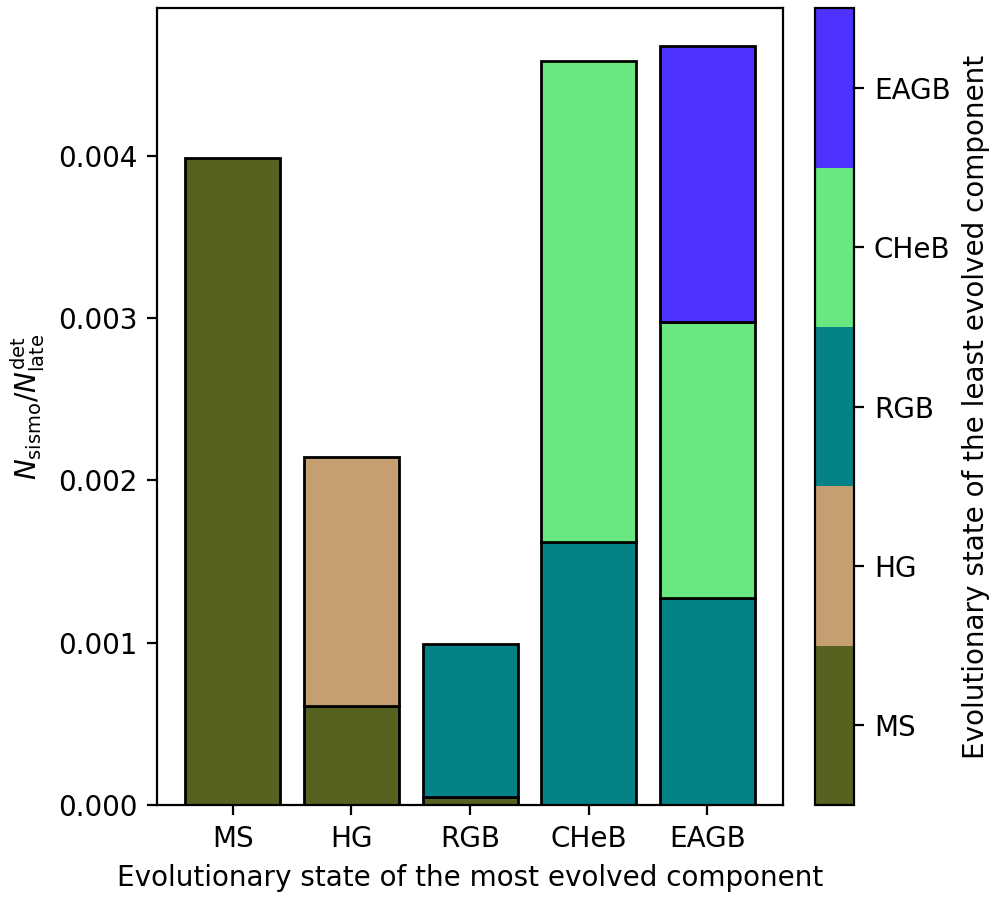}
    \caption{Histogram of the counts of the different types of asteroseismic binaries (N$_\text{seismo}$) produced by the simulation that adopts the MDS17 initial parameter distribution relative to the number of single stars with detectable oscillations in the evolutionary phase of the most evolved star of the system (N$_\text{late}^\text{det}$).}
    \label{fig:hist_detectable_bin}
\end{figure}

Although the \cite{chaplin11} model is capable of calculating the detection probability for each component of a binary, it only considers their individual backgrounds and asteroseismic signals, as if each star was photometrically resolved.
In the case of a binary system that is unresolved by \Kepler{}, the probability of detecting the asteroseismic signal of both stars in the combined power spectrum has an additional dependence on the background and asteroseismic signal of the companion.

The model we use to compute the probability of detecting oscillations in the components of a binary is the one presented in \cite{miglio14} and returns the detection probability of the primary $p_{\text{seismo}, \text{pri}}$ and secondary $p_{\text{seismo}, \text{sec}}$ stars in the combined power spectrum, accounting for the dilution of the seismic signal due to the presence of the companion. Then, the probability of detecting oscillations for both components of the binary is computed as
\begin{equation}
    p_{\text{seismo, bin}} = p_{\text{seismo, pri}} \times p_{\text{seismo, sec}} \, .
    \label{eq:detection_prob}
\end{equation}
Two stars with a 90\% probability each of having oscillations detected result in an overall probability to detect both equal to $p_{\text{seismo, bin}}=81\%$. However, we choose to impose a slightly more stringent condition for a binary star to be likely identified as an asteroseismic binary and require $p_{\text{seismo, bin}} \geq 90\%$.

It should be mentioned that the method we adopt ignores effects that decrease the amplitude of modes apart from the dilution of asteroseismic signal due to the presence of a companion\footnote{ In an unresolved binary system, the oscillations of a star are observed as a fluctuation of the total flux of the binary, therefore the relative amplitude of the signal will be smaller than in the case of a single star. We note, however, that the same phenomenon can occur with contamination from a bright neighbor or in the case of a chance alignment.}.
For instance, suppression of the amplitudes of observed modes has been found for red giants in short-period binaries, with modes being completely undetected in some of the systems with periods below 60 days.
The mechanism is currently believed to be due to tidally induced magnetic activity \citep{gaulme14, benbakoura21}.
Although we lack clear prescriptions to simulate this dampening, we can check how many systems would be close enough to show a significant reduction in the amplitude of the modes.
Selecting systems with a sum of radii larger than $16\%$ of the orbital semi-major axis, the largest ratio for which \cite{gaulme14} notes that binaries always show modes of oscillations, we find one double MS, two double RGBs and three double CHeB.
These represent the 0.002\%, 0.8\% and 4.6\% of the total number of their respective type of binaries.
Finally, in our analysis we also have not investigated the effect of light contamination due to blending or chance alignments, which will need to be carefully assessed in the real asteroseismic data to avoid false positives, both in single and in binary stars.

From Table~\ref{tab:counts_phases_det}, about $47.3\%$ of the RGBs and $68.6\%$ of the CHeBs in binaries have detectable oscillations.
These fractions are smaller than for the case of single stars presented in Section~\ref{sec:detection-sin}. This is mainly caused by the dilution of light due to the presence of a companion discussed above.

The total number of red giant stars with detectable oscillations, comprising RGB, CHeB and early-asymptotic giant branch (EAGB), is $41537$, regardless of their binary status, while those in binaries are $11544$, and $227$ are in asteroseismic binaries.
For every $1000$ red giant stars (RGB, CHeB and EAGB) with detectable oscillations, about $140$ are in binary systems and $2.7$ are in asteroseismic binaries.

Table~\ref{tab:counts_binaries} presents a breakdown of the types of binary system produced in the simulation, with components up to the EAGB phase.
Of 257 double RGB and 145 mixed CHeB and RGB systems, 153 ($\sim 60\%$) and 98 ($\sim 68\%$) respectively have at least one component with detectable oscillations.
In the case of double CHeB stars, this happens for 55 of the 65 systems ($\sim 85\%$) produced by the simulation.
Overall, in more than half of the binary systems hosting two giant stars the detection of an asteroseismic signal for at least one of the stars should be possible.

The number of asteroseismic binaries is dominated by double CHeB binaries ($35.1\%$), as, without interaction between the components, this phase has a long lifetime, followed by binaries with a CHeB and an RGB ($19.2\%$) and double RGBs ($13.2\%$).
Asteroseismic binaries containing stars in the EAGB phase are less common ($7.3\%)$, while those with MS or HG stars contribute a relatively high fraction ($24.5\%$).

Double CHeB binaries also have the largest fraction of systems detected as asteroseismic binaries, namely $82\%$, which is expected since the components of these binaries should have very similar mass, radius, and luminosity.
Double RGB systems can be detected as asteroseismic binaries in $7.8\%$ of the cases, and mixed RGB and CHeB in $20\%$ of the cases.
Mixed EAGB binaries, although limited in number, show a moderate fraction ($38.9\%$) of asteroseismic binaries, and double EAGB are all asteroseismic binaries.

We can compare these results to the counts of stars with detectable oscillations in different evolutionary phases.
For every $1000$ RGB stars with detectable oscillations, there are $0.9$ double RGB asteroseismic binary systems, $1.4$ composed of a CHeB and an RGB, and $0.1$ with an EAGB and an RGB.
Analogously, for every 1000 CHeB stars with detectable oscillations there are $3.0$ asteroseismic double CHeB binaries and $1.6$ with a CHeB and and RGB.

In Figure~\ref{fig:hrd_detectable_bin} we show a Hertzsprung-Russell diagram (HRD) of single and binary stars with detectable oscillations, the latter all considered unresolved, and selecting only stars in evolutionary phases between the MS and the EAGB.
Figure~\ref{fig:hist_detectable_bin} instead presents the fraction of each type of asteroseismic binary with respect to single stars with detectable oscillations.

\begin{figure}[t]
        \centering
        \includegraphics[width=\linewidth]{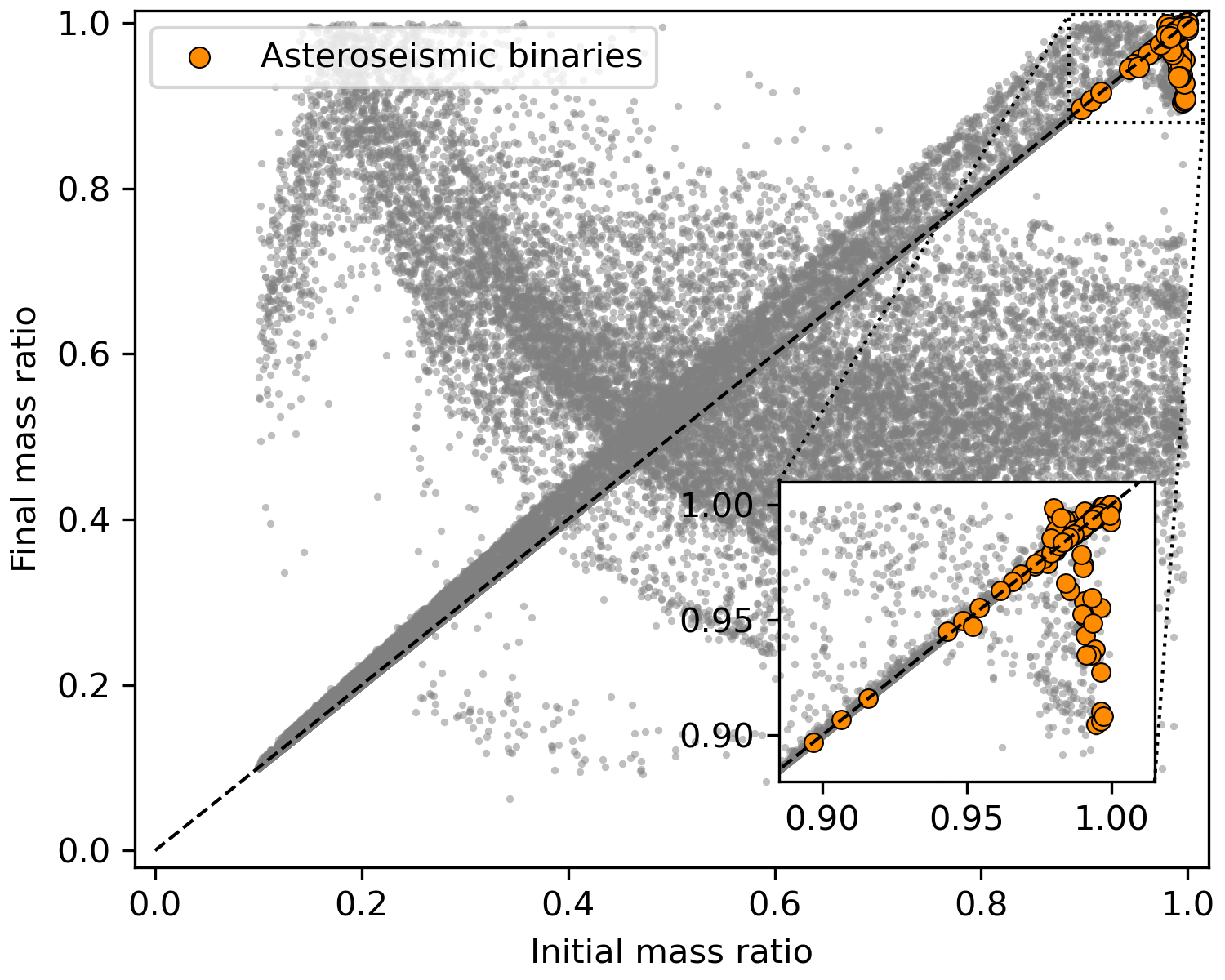}
        \caption{Initial and final mass ratio of the binary stars in the MDS17 simulation. The gray points represent all binaries in the simulation, and the colored circles mark systems that can be detected as asteroseismic binaries. Systems near the black dashed line have almost identical initial and final mass ratio. The inset highlights the asteroseismic binaries.}
        \label{fig:mass_ratio}
    \hfill
\end{figure}

\begin{figure}[t]
    \centering
        \includegraphics[width=\linewidth]{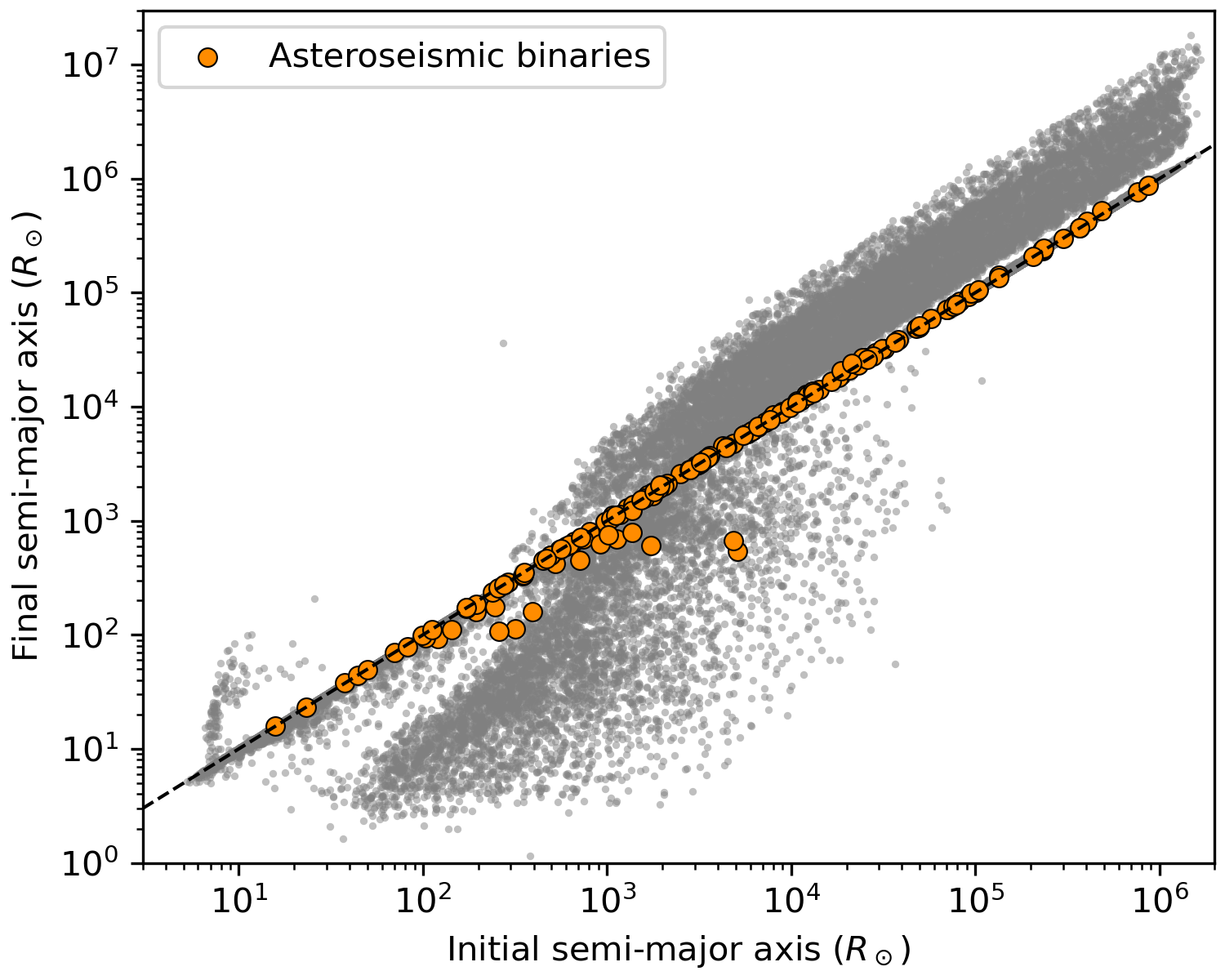}
        \caption{Same as Figure~\ref{fig:mass_ratio}, but showing initial and final semi-major axis of the binary systems. A semi-major axis of $\SI[parse-numbers=false]{10^3}{\Rsun}$ typically corresponds to a period of around $\SI{3000}{\day}$ in the simulation.}
        \label{fig:semimajoraxis}
\end{figure}

\subsection{Orbital properties of asteroseismic binaries}
The initial and final mass ratios of the binaries produced by the simulation are shown in Figure~\ref{fig:mass_ratio}.
The asteroseismic binaries (orange circles) are characterized by initial and final mass ratios close to unity, but appear to split in two groups.
The first group displays almost no change in the mass ratio, and is unlikely to have undergone any significant mass exchange during the evolution.
The second group has initial mass ratios very close to $1$, while the final ones span an interval between $0.9$ and $1$, with changes up to $10\%$ from their initial values.
Additionally, their semi-major axis changes by $\lesssim 5\%$ from its initial value and their final eccentricity is almost identical to the initial one.
We can explain the change in the mass ratio of this second group of asteroseismic binaries just with a Reimers' mass loss \citep{reimers1975, reimers1977}, which TRIEGAL adopts with $\eta=0.3$ for the RGB. At the end of the simulation, the initially slightly more massive component of the binary system has already moved to the CHeB, losing some mass while ascending the RGB before eventually igniting He in the core, while its less massive companion is still on the RGB or in earlier phases of the evolution and has not yet experienced a similar mass loss.

Figure~\ref{fig:semimajoraxis} shows the initial and final semi-major axis of all the binaries in the simulation, with the asteroseismic binaries highlighted in the same way as in Fig.~\ref{fig:mass_ratio}.
In this case, we can also distinguish two groups of asteroseismic binaries.
The first, most populated group, shows a final semi-major axis almost identical to the initial one, while the second has more significant variations. In this second group, the semi-major axis of all systems decreased during the evolution of the binary.
These binaries have not changed their mass ratio appreciably and, therefore, did not undergo any relevant mass transfer process, but have started the simulation with a high initial eccentricity and ended fully circularized.
Tidal interaction must have occurred between the components, leading to circularization of their orbits \citep{pricewhelan18_circularization}.
Additionally, we note that this interaction should leave an imprint in the rotation of the stars, in the form of a spin-up of stellar rotation \citep{gaulme20}.

The conclusion we can draw from the analysis presented above is that we can detect as asteroseismic binaries those systems where the components (a) were born with mass ratio close to one and (b) during their evolution did not have the chance to exchange mass, keeping their mass ratio and luminosity ratio both close to one.
It seems unlikely that systems that interacted significantly during their evolution can be detected as asteroseismic binaries, however, this does not preclude the possibility that one of the components still has detectable oscillations.

\subsection{Double CHeB binaries}

\begin{figure}
    \centering
    \includegraphics[width=\columnwidth]{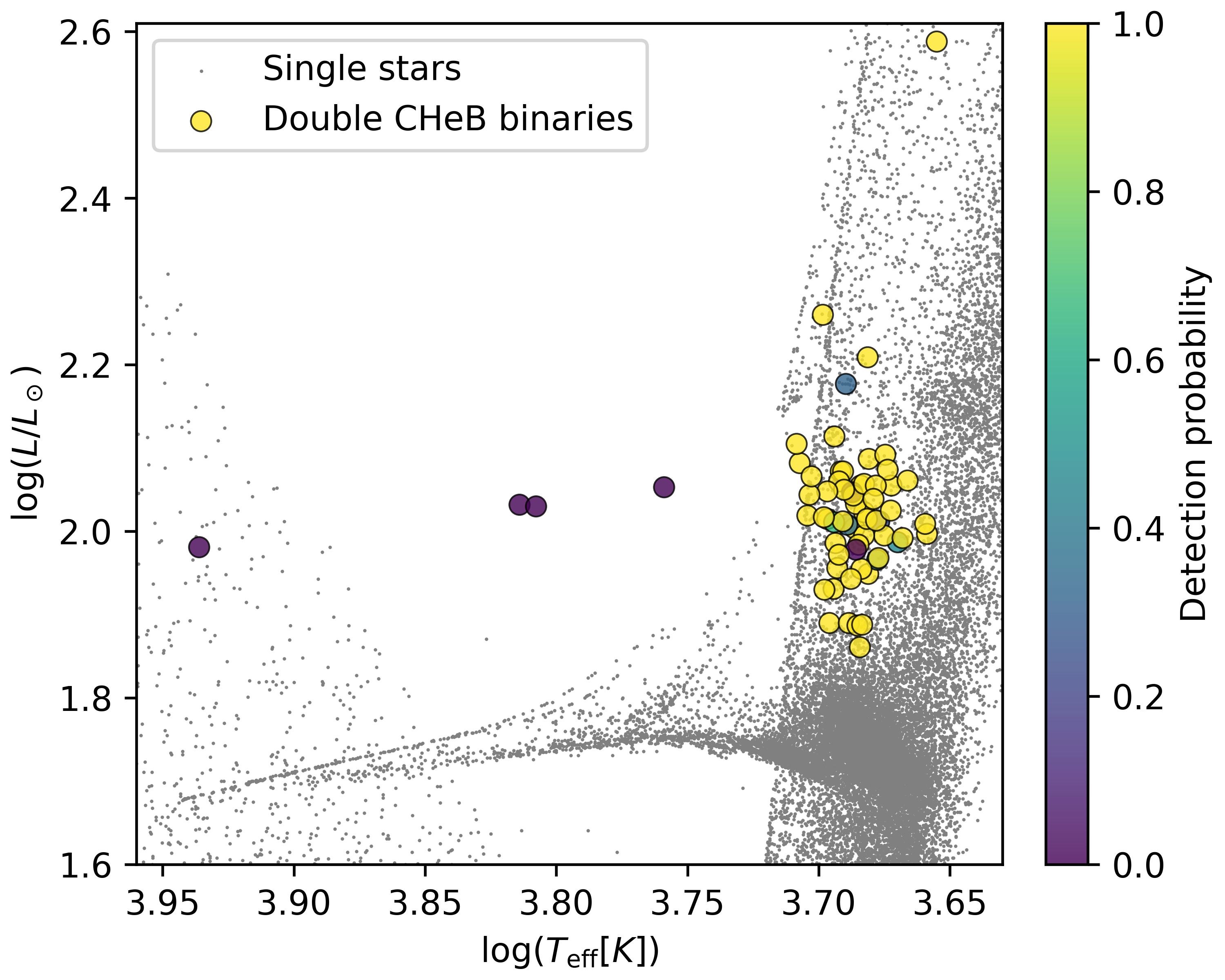}
    \caption{Location on the HRD of the double CHeB stars produced by the MDS17 simulation. Small gray dots represent single stars, while colored points represent the double CHeB stars, with the color indicating the detection probability.}
    \label{fig:double_rc}
\end{figure}

\begin{figure}
    \centering
    \includegraphics[width=\columnwidth]{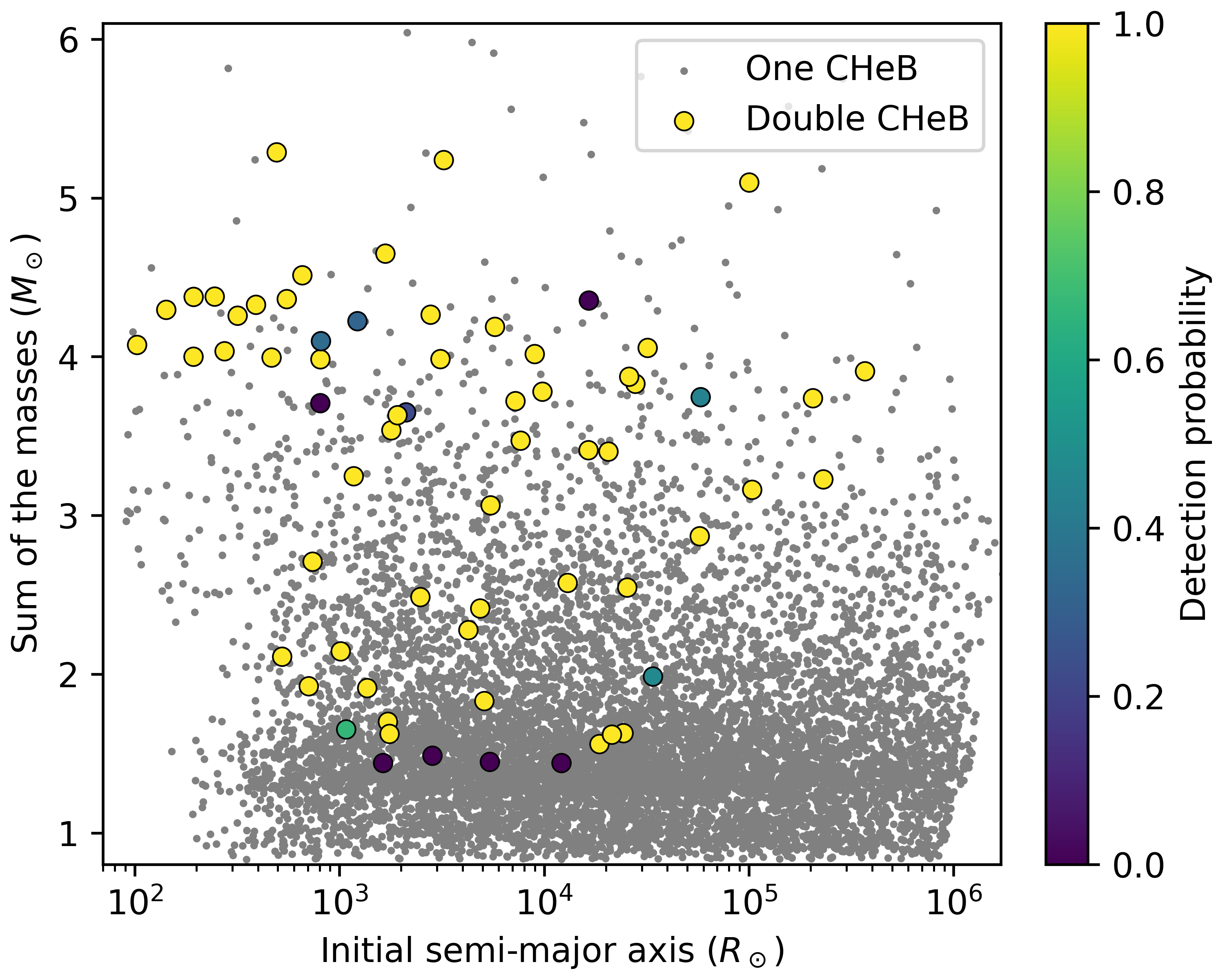}
    \caption{Initial semi-major axis and total mass at the end of the MDS17 simulation ($M_1 + M_2$) for binaries containing of one (gray points) and two (colored circles) CHeB stars. The color scale represents the probability to observe each double CHeB system as an asteroseimic binary.}
    \label{fig:double_rc_sma_mtot}
\end{figure}

The evolution of a binary system depends not only on its initial orbital binary parameters, but also on the properties of the individual components.
In fact, even without interactions, in a system where the two components have significantly different masses, the more massive one rapidly moves to later phases of the evolution.
For double CHeB binaries, this means that their components must have been born with similar masses, and must have evolved without experiencing significant mass exchange or mass loss until the ignition of He in the core.

In Fig.~\ref{fig:double_rc} we show the position on an HRD of the double CHeB binaries produced by the simulation using their unresolved luminosity, which is roughly twice that of a single red clump (RC) star, and we also note that most of these binaries have a mass ratio close to one.
In most of these systems, the oscillations for both components are expected to be detectable, and consequently a large fraction of double CHeB binaries are expected to be asteroseismic binaries.
Regarding the systems that have a low probability of being detected as asteroseismic binaries, most of them are faint, with magnitudes in \Kepler{}'s band larger than 14.
In addition to being faint, four systems have components that are hotter than or close to the temperature of the red edge of the classical instability strip, as defined in \cite{chaplin11}, and two have a luminosity ratio close to $0.5$.

Figure~\ref{fig:double_rc_sma_mtot} shows the double CHeB binaries in the MDS17 simulation in terms of their initial semi-major axis and the total mass of the system at the end of the simulation.
In particular, there are no double CHeB stars with an initial semi-major axis below $\SI{500}{\Rsun}$ and a total mass at the end of the simulation smaller than $\SI{4}{\Msun}$, while there are double CHeB systems in smaller orbits with a final total mass larger than $\SI{4}{\Msun}$.
The former region of the semi-major axis - mass plane should be populated by double RC binaries, and their absence from this part of the parameter space must be a consequence of previous binary interactions.
In fact, when both components of a binary system are low-mass stars ($M\lesssim \SI{2}{\Msun}$), they ascend the RGB until the tip before finally undergoing the He flash and then quiescently burning He in the core in the RC.
Close to the tip, their radii can reach $\sim\SI{150}{\Rsun}$ depending on their mass and metallicity, as shown in stellar tracks (e.g. those from \citealt{nguyen22_parsecV2}).

Approximating the Roche lobe of a star with a sphere that has the same volume, the radius $r_1$ of said sphere can be computed using eq. 4 of \cite{paczynski1971} in the case of $q=M_1/M_2 \approx 1$
\begin{equation}\label{eq:roche-lobe}
    \frac{r_1}{A} = 0.38 + 0.2 \log \left(q\right) \approx 0.38
\end{equation}
where $A$ is the orbital separation of the binary.
For a binary with orbital separation $\leq\SI{500}{\Rsun}$, this leads to $r_1 \lesssim \SI{150}{\Rsun}$.
Therefore, binaries composed of two low-mass stars with a separation approaching this threshold will likely experience Roche-lobe overflow of the primary leading to mass transfer, causing a divergence of the stellar evolution of both stars which does not result in an observable double RC binary.
If the components of the binary have slightly larger masses, $\gtrsim \SI{2}{\Msun}$ dependent on metallicity, they will ignite He in the core without reaching the RGB tip, at smaller radii, and continue their evolution as a double secondary clump binary.
This explains why in Fig.~\ref{fig:double_rc_sma_mtot} several double CHeB systems with total mass larger than \SI{4}{\Msun} have a semi-major axis smaller than \SI{500}{\Rsun}.
A real example of this kind of system is KIC~9246715 \citep{helminiak15, rawls16}, an eclipsing binary hosting two secondary clump stars with masses $M_1=2.171^{+0.006}_{-0.008}\,\si{\Msun}$ and $M_2=2.149^{+0.006}_{-0.008}\,\si{\Msun}$, for a total mass of about \SI{4.32}{\Msun}.
These stars had a radius of approximately \SI{25}{\Rsun} at the tip of the RGB, which is substantially smaller than their expected Roche lobe $r_1 \approx \SI{80}{\Msun}$ (according to Eq.~\ref{eq:roche-lobe} with a mass ratio similar to their present one), therefore allowing the system to survive as a double CHeB binary.
Although this is an example of a binary composed of two secondary clump stars, \cite{rawls16} found clear evidence of oscillation modes from only one component, but the oscillation modes were wider than typically expected from similar stars and could be due to mode overlap in the power spectrum.

Double RC binaries therefore constitute one of the few subsets of asteroseismic binaries where clear and well-defined limitations are imposed from the interplay of stellar evolution and binary interaction theory during their past, and their study can aid in placing crucial constraints on aspects of both theories that are still uncertain.
Requiring that an observed double RC binary survived an interaction during the ascent of both components along the RGB should then lead to strong constraints on the orbital configuration (semi-major axis and eccentricity), dependent on the expected RGB-tip radius and surface gravity.
Furthermore, as was shown in Sect.~\ref{sec:seis_bin}, even for less extreme mass transfer events where the binary survives, it is not expected to be detectable as an asteroseismic binary.

\subsection{Under-massive stars}
\label{sec:results-under}

\begin{table*}[t]
    \caption{Counts of under-massive stars in the simulation adopting the \citet{moedistefano2017} prescription, for evolutionary phases from MS to EAGB.}
    \label{tab:counts_under-massive}
    \centering
    \begin{tabular}{lrrrrrrrr}
        \toprule
        Phase &
        N &
        N$_\text{lone pri}$ &
        N$_\text{sec}$ &
        N$_\text{lone pri}^\text{det}$ &
        N$_\text{sec}^\text{det}$ &
        N$_\text{lone pri}^\text{det}$ / N$^\text{det}$ &
        N$_\text{sec}^\text{det}$ / N$^\text{det}$ &
        N$^\text{det}$ \\
        \midrule
        MS & 3 & 0 & 3 & 0 & 0 & 0.00000 & 0.00000 & 7528 \\
        HG & 5 & 0 & 5 & 0 & 1 & 0.00000 & 0.00031 & 3262 \\
        RGB & 34 & 30 & 4 & 28 & 2 & 0.00132 & 0.00009 & 21227 \\
        CHeB & 290 & 287 & 3 & 279 & 1 & 0.01558 & 0.00006 & 17906 \\
        EAGB & 21 & 21 & 0 & 20 & 0 & 0.00850 & 0.00000 & 2354 \\
        \bottomrule
    \end{tabular}
    \tablefoot{
        Column names including the term "det" refer to the subsample of stars with detectable oscillations. The last column is for reference and is the same as in Table~\ref{tab:counts_phases_det}. There are no under-massive primary stars still in binaries, therefore the corresponding columns are not reported in the Table.
    }
\end{table*}

\begin{figure}[t]
    \centering
    \includegraphics[width=\columnwidth]{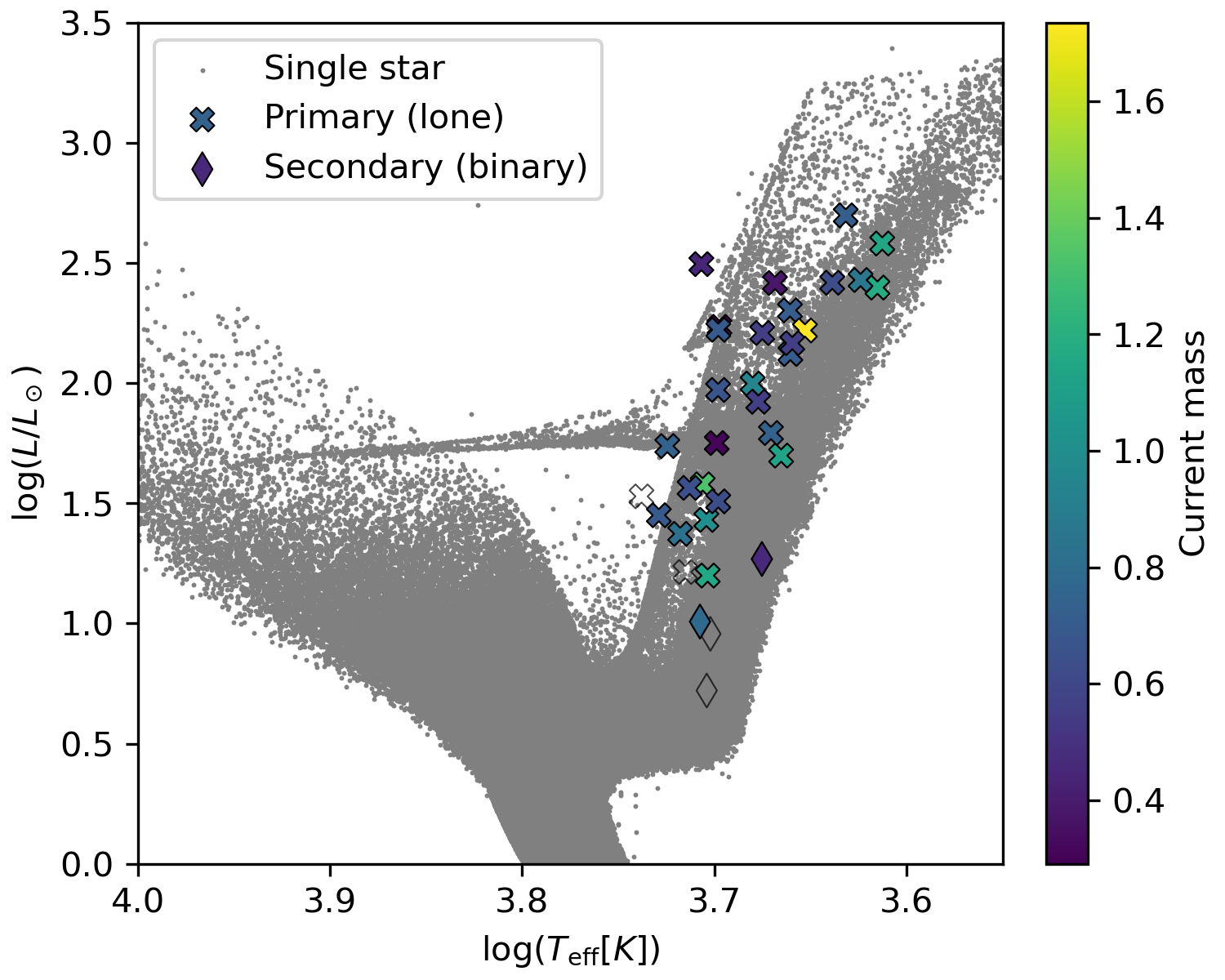}
    \caption{Location on an HRD of under-massive stars in the RGB phase for the MDS17 simulation. Crosses indicate under-massive lone primaries while thin diamonds mark under-massive secondary stars in a surviving binary. The color represents the current mass of each under-massive star, and empty markers indicate the ones where detection of oscillations is not likely ($p_\text{seismo,bin}<0.9$). Gray points represent single stars from the simulation and are used to draw the main loci of the HRD.}
    \label{fig:hrd_under-massive_rgb}
\end{figure}

\begin{figure}[t]
    \centering
    \includegraphics[width=\columnwidth]{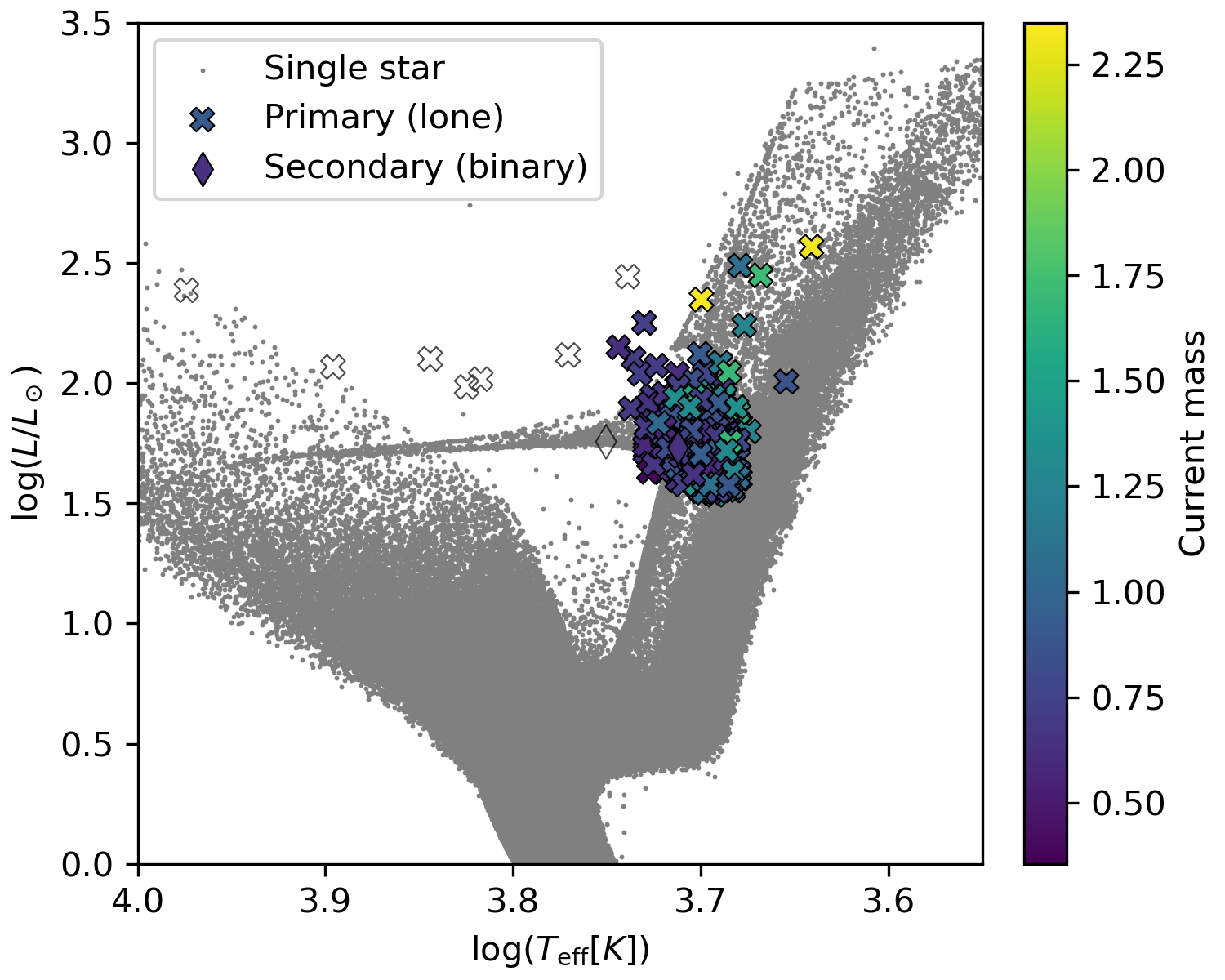}
    \caption{Same as Fig.~\ref{fig:hrd_under-massive_rgb}, but for under-massive stars in the CHeB phase.}
    \label{fig:hrd_under-massive_cheb}
\end{figure}

According to stellar evolution theory, single stars continuously lose mass throughout their life.
During the MS, the amount of mass lost by low-mass stars is small, and in the case of the Sun it is estimated to be of the order of $\SI{e-14}{\Msun\per\yr}$.
Subsequent phases of the evolution, however, can lead to a more significant loss of mass, and the RGB phase is particularly important in this context.
The total mass lost by a star during the RGB phase is still a matter of debate, but ensemble averages for \Kepler{} field stars \citep[see for example][]{yu21, miglio21, Brogaard24} yield $\Delta M_\text{RGB} \approx \SI{0.1}{Msun}$, with a dependence on metallicity.

Asteroseismic studies performed on members of stellar clusters have nonetheless highlighted that there exists a population of stars with low masses that do not match their evolutionary phase given the age of the cluster itself and would require high to extremely high mass losses to fit in the context of single stellar evolution.
For example, \cite{anthonytwarog13} found that the star W007017 (KIC~4937011) in the open cluster NGC~6819 is a Li-rich star and \cite{handberg17} determined that it is a RC star with a mass of $\SI{0.71 \pm 0.08}{\Msun}$, much lower than the other RC stars in the cluster.
\cite{brogaard21_ngc6791} investigated eleven giant stars, members of the old open cluster NGC 6791, and discovered that KIC 2436543 is an RC star with a mass of $M=\SI{0.90 \pm 0.07}{Msun}$, again lower than any other RC stars in the cluster.
\cite{matteuzzi23} also presented an analysis of \Kepler{}'s observations and identified 11 RC stars in the field with masses below $\sim\SI{0.8}{\Msun}$ and coupling factors between the pressure and gravity resonance cavities much higher than usual RC stars, although some of these, being metal poor, are most likely red Horizontal Branch stars.
Some of these stars are likely the result of a binary interaction, where one or more mass transfer events occurred and ultimately led to the disruption of one of the components of the binary system. For instance, \cite{matteuzzi24} investigated in detail KIC~4937011 and constrained its formation scenario as a direct consequence of a common envelope phase involving an RGB star and a MS star.

On this premise, it is worthwhile to search for such stars in the simulation.
The first step consists of defining a criterion to assign the ``under-massive`` label to a star.
In the simulation, single stars with initial masses $M \gtrsim \SI{1.7}{\Msun}$ generally lose a very small fraction of their mass during their evolution.
Stars with a smaller mass ($M \lesssim \SI{1.7}{\Msun}$) instead, as a consequence of having to climb the RGB until its tip to undergo the He-flash, can lose on average \SI{0.1}{\Msun}.

To ensure we are able to distinguish single-star mass loss from binary mass interaction, we ultimately choose to mark as under-massive the stars that at the end of the simulation have lost $20\%$ or more of their initial mass.
Table~\ref{tab:counts_under-massive} provides a summary of the counts of these stars for evolutionary phases from the MS to the EAGB.
Most under-massive stars are in the CHeB phase and are lone primaries, like KIC 4937011, with only a few being under-massive secondary stars.
Overall, a large fraction of the evolved (i.e. in the RGB and later phases) under-massive stars have a high probability of having oscillations detected.
For every 1000 RGB stars with detectable oscillations, regardless of their binary state, $1.4$ are under-massive. In the case of CHeB stars, instead, for every 1000 with detectable oscillations, $15.6$ are under-massive.

Figures~\ref{fig:hrd_under-massive_rgb} and \ref{fig:hrd_under-massive_cheb} present the location on the HRD of the under-massive stars in the RGB and CHeB phases, respectively, and mark with empty symbols those whose oscillations are unlikely to be detected.
In the RGB, the under-massive secondary stars appear in the low-luminosity part of the branch, while the lone primaries are located at higher luminosities, with a small overlap around $\log L \gtrsim 1.2$.
Under-massive CHeB stars, instead, are located mostly around the RC, with some showing higher temperatures and appearing in locations of the HRD normally devoid of stars that are unlikely to have oscillations detected if their temperature is higher than the red edge of the classical instability strip.
Finally, a few under-massive CHeB stars have larger luminosities and are slightly cooler than the RC.

\begin{figure}
    \centering
    \includegraphics[width=\columnwidth]{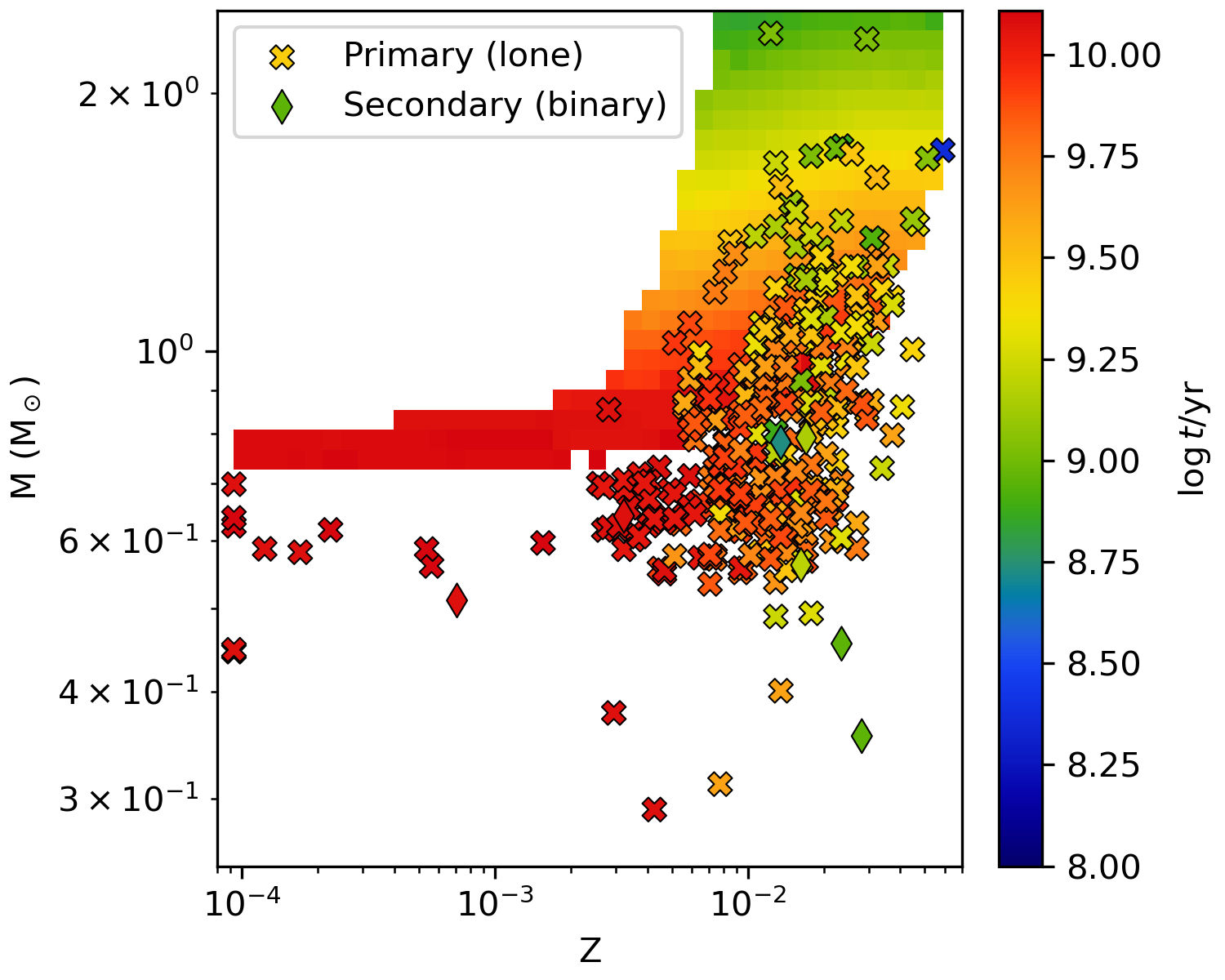}
    \caption{Distribution of mass versus metallicity for under-massive giant stars (RGB, CHeB and EAGB), with age indicated by the color. The 2D histogram is made using single giant stars and the color represents the median age of stars in each bin.}
    \label{fig:Z_mass_under}
\end{figure}

Finally, Figure~\ref{fig:Z_mass_under} presents the masses and metallicities of just the under-massive giant stars with detectable oscillations on top of a 2D histogram computed from the masses and metallicities of the stars born single, with the color describing the median age in each mass-metallicity bin.
Most of the under-massive stars currently have a combination of mass and metallicity such that they land on a region of the 2D histogram where the median age of the single stars is older than their actual age, thus making them look older from an observational point of view if not properly identified.

\subsection{Over-massive stars}
\label{sec:results-over}

\begin{table*}[t]
    \caption{Same as Table~\ref{tab:counts_under-massive}, but for the case of over-massive stars.}
    \label{tab:counts_over-massive}
    \centering
    \begin{tabular}{lrrrrrrrrrrr}
        \toprule
        Phase &
        N &
        N$_\text{lone pri}$ &
        N$_\text{pri}$  &
        N$_\text{sec}$  &
        N$_\text{lone pri}^\text{det}$ &
        N$_\text{lone pri}^\text{det}$ / N$^\text{det}$ &
        N$^\text{det}$ \\
        \midrule
        MS & 41 & 6 & 20 & 15 & 0 & 0.00000 & 7528 \\
        HG & 12 & 4 & 8 & 0 & 1 & 0.00031 & 3262 \\
        RGB & 57 & 53 & 4 & 0 & 45 & 0.00212 & 21227 \\
        CHeB & 236 & 236 & 0 & 0 & 226 & 0.01262 & 17906 \\
        EAGB & 22 & 22 & 0 & 0 & 22 & 0.00935 & 2354 \\
        \bottomrule
    \end{tabular}
    \tablefoot{There are no over-massive primary or secondary stars, i.e. still in a binary system, with detectable oscillations, therefore their respective columns have not been included in the Table.}
\end{table*}

\begin{figure}[t]
    \centering
    \includegraphics[width=0.98\columnwidth]{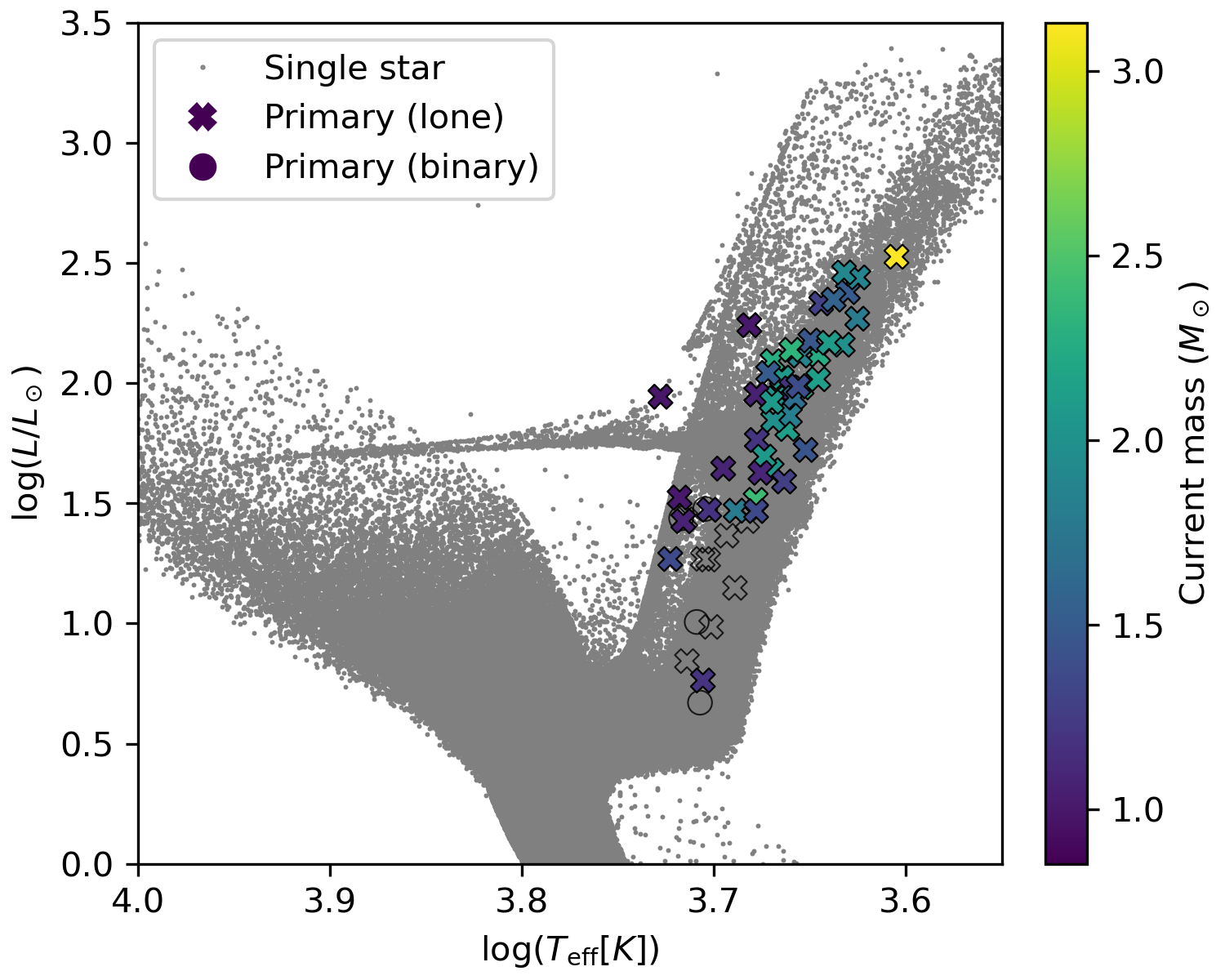}
    \caption{HRD of the single stars (gray points) with plotted on top the location of the over-massive stars in the RGB phase. The circles diamonds mark primary stars in a binary that are over-massive, while crosses indicate the over-massive lone primaries. The color represents the current mass of each over-massive star, and empty markers indicate the ones where detection of oscillations is not likely (P<0.9).}
    \label{fig:hrd_over-massive_rgb}
\end{figure}

\begin{figure}
    \centering
    \includegraphics[width=0.98\columnwidth]{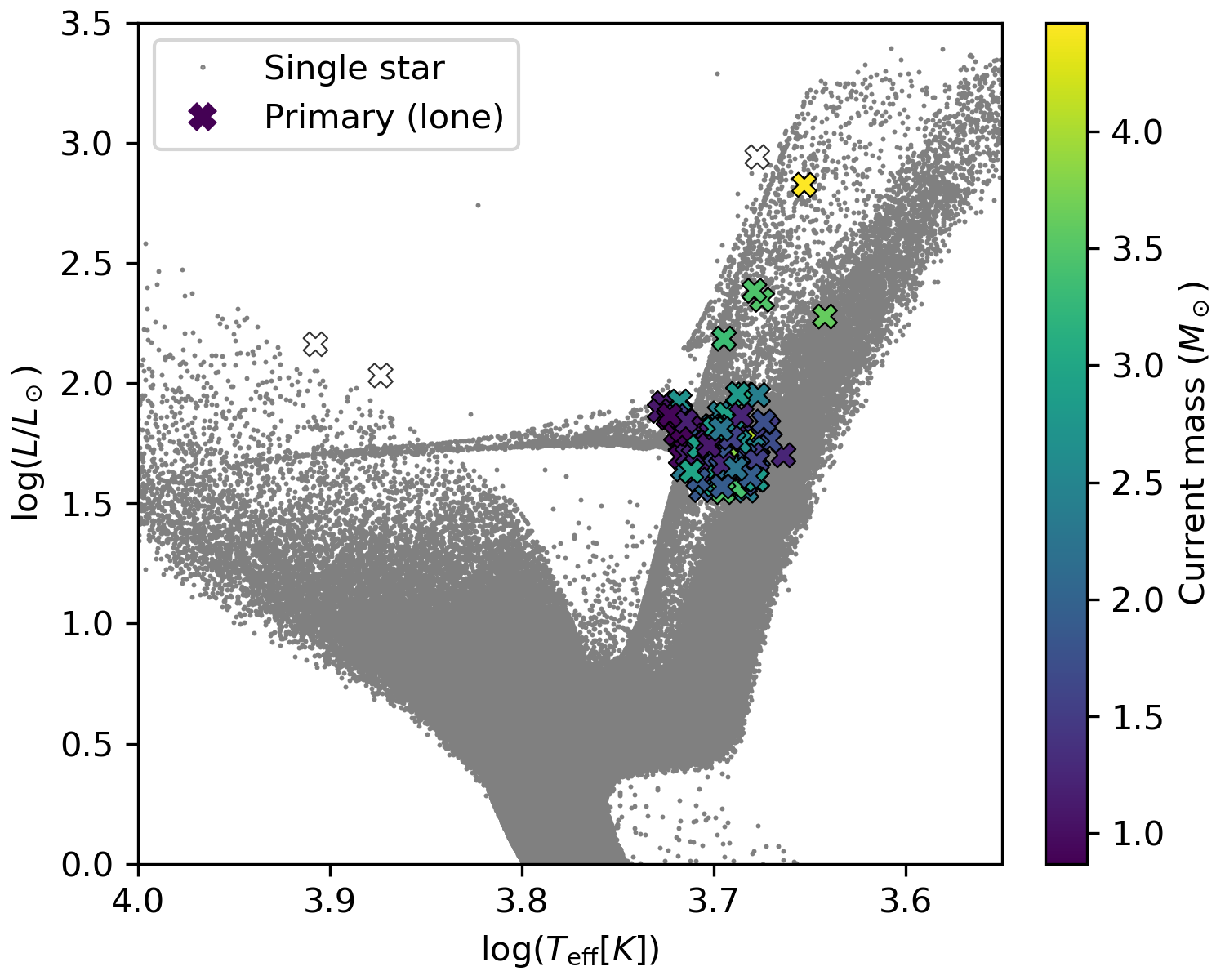}
    \caption{Same as Fig.~\ref{fig:hrd_over-massive_rgb}, but for over-massive stars in the CHeB phase.}
    \label{fig:hrd_over-massive_cheb}
\end{figure}

In the context of single stellar evolution, stars are not expected to gain mass at any stage of their life.
However, studies of stellar clusters \citep[see for instance][]{brogaard16, leiner16, handberg17, brogaard21_ngc6791} have identified cluster members that have a mass substantially larger than expected given their evolutionary phase and the age of the cluster itself.
These stars are referred to as ``over-massive stars'' and can be explained as the result of a mass exchange in a binary system.

When observing field stars, it is more difficult to identify over-massive stars based only on their mass and evolutionary stage, but chemistry provides a way to tag them.
For instance, the age estimated for the young $\alpha$-rich (YAR) stars, a group of over-massive stars with chemical composition very similar to that of the high-$\alpha$ sequence, excludes that they are part of the old population \citep{chiappini15, martig15, izzard18} unless their current masses are higher than the initial ones.

Similarly to the case of under-massive stars, we select over-massive stars from the simulation using a threshold of 1\% of mass gained at the end of the simulation.
This approach leads to some differences with the literature, where over-massive stars are generally more massive than the typical RGB star.
For example, our definition includes low-mass stars that acquire additional mass or undergo mergers in the early phases of evolution, ultimately exhibiting evolutionary patterns similar to those of a standard RGB star.

Table~\ref{tab:counts_over-massive} shows the counts of the over-massive stars from the MS to the EAGB, while Figure~\ref{fig:hrd_over-massive_rgb} and \ref{fig:hrd_over-massive_cheb} show the location on the HRD of stars in the RGB and CHeB phases, respectively.
Most of these are lone primaries, with just a couple of over-massive primaries in the RGB phase.
For every 1000 RGB stars with detectable oscillations, regardless of their binary state, $2.1$ are over-massive, while the over-massive are $12.6$ for every 1000 CHeB stars with detectable oscillations.

\begin{figure}
    \centering
    \includegraphics[width=\columnwidth]{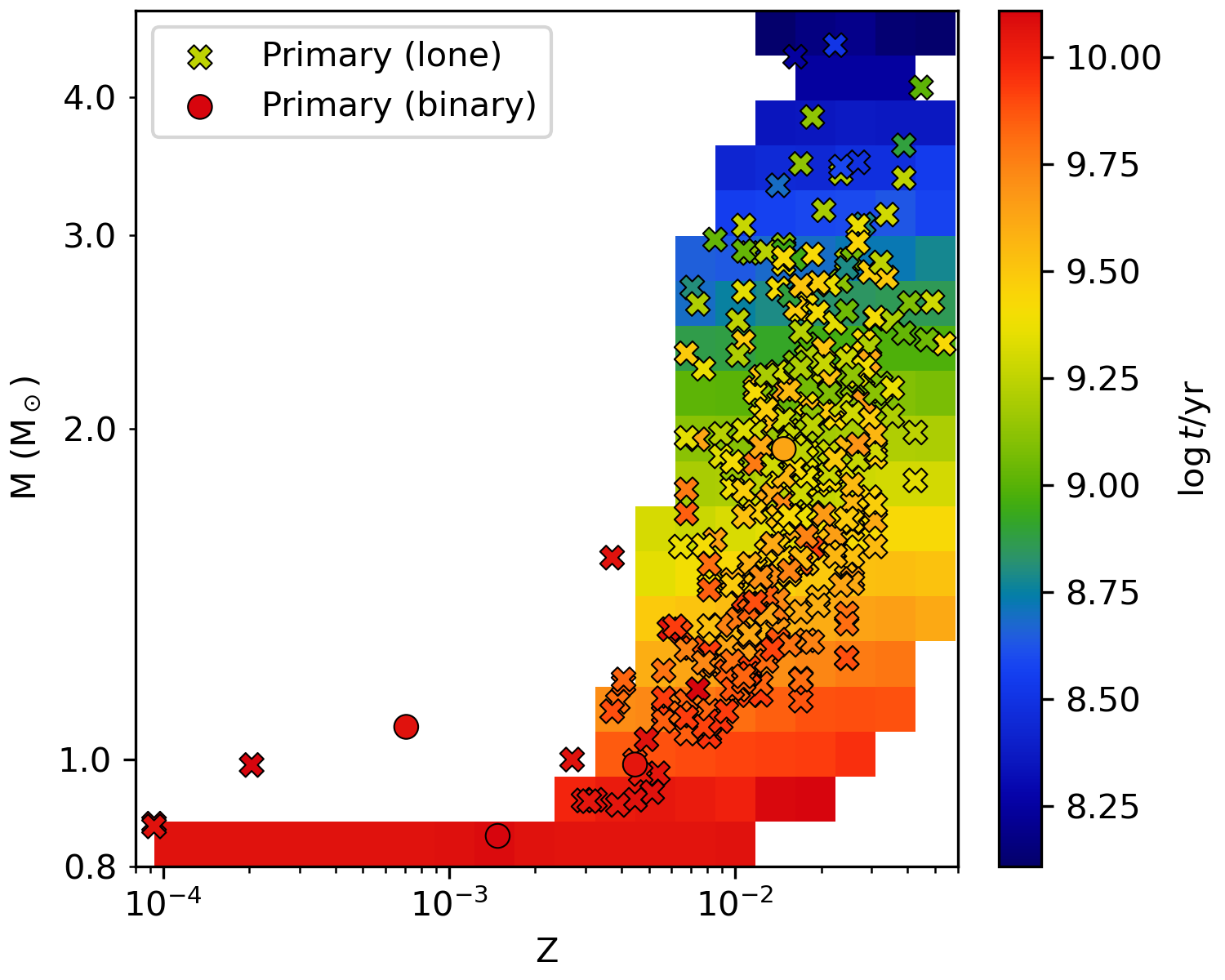}
    \caption{Similar to Figure~\ref{fig:Z_mass_under}, but for over-massive stars.}
    \label{fig:Z_mass_over}
\end{figure}

Looking at the mass vs metallicity distribution of the over-massive giant stars in Figure~\ref{fig:Z_mass_over}, they generally fall in cells of the 2D distribution of the single giant stars that are characterized by smaller median ages.
That is, these stars would appear younger than they actually are if treated as single stars.

\section{Comparison with other prescriptions}
\label{sec:comparison}

\begin{table*}[t]
    \caption{Same as Table~\ref{tab:counts_binaries}, but for binaries with detectable oscillation produced by the simulation with the E06 and non-interacting binary prescriptions.}
    \label{tab:counts_binaries_compare}
    \centering
    \begin{tabular}{llrrrrrrr}
        \toprule
        phase$_\text{late}$ & phase$_\text{early}$ & N & N$_\text{one}$ & N$_\text{seismo}$ & N$_\text{seismo}$ / N$_\text{late}^\text{det}$ & N$_\text{seismo}$  / N$_\text{early}^\text{det}$ & N$_\text{seismo}$ / $\sum$ N$_\text{late}^\text{det}$ & N$_\text{seismo}$/$\sum$N$_\text{seismo}$ \\
        \midrule
        \multicolumn{9}{c}{E06}\\
        \midrule
        MS & MS & 55691 & 1908 & 22 & 0.00287 & 0.00287 & 0.00041 & 0.39286 \\
        HG & MS & 7275 & 913 & 1 & 0.00030 & 0.00013 & 0.00002 & 0.01786 \\
         & HG & 253 & 6 & 6 & 0.00181 & 0.00181 & 0.00011 & 0.10714 \\
        RGB & RGB & 87 & 38 & 7 & 0.00032 & 0.00032 & 0.00013 & 0.12500 \\
        CHeB & RGB & 31 & 15 & 7 & 0.00039 & 0.00032 & 0.00013 & 0.12500 \\
         & CHeB & 12 & 0 & 11 & 0.00062 & 0.00062 & 0.00021 & 0.19643 \\
        EAGB & CHeB & 3 & 1 & 2 & 0.00082 & 0.00011 & 0.00004 & 0.03571 \\
        \midrule
        \multicolumn{9}{c}{Non-interacting}\\
        \midrule
        MS & MS & 62384 & 1150 & 86 & 0.01199 & 0.01199 & 0.00147 & 0.26462 \\
        HG & MS & 6899 & 599 & 2 & 0.00054 & 0.00028 & 0.00003 & 0.00615 \\
         & HG & 684 & 23 & 12 & 0.00325 & 0.00325 & 0.00021 & 0.03692 \\
        RGB & MS & 7953 & 4947 & 1 & 0.00004 & 0.00014 & 0.00002 & 0.00308 \\
         & HG & 998 & 523 & 1 & 0.00004 & 0.00027 & 0.00002 & 0.00308 \\
         & RGB & 351 & 202 & 30 & 0.00124 & 0.00124 & 0.00051 & 0.09231 \\
        CHeB & HG & 385 & 376 & 1 & 0.00005 & 0.00027 & 0.00002 & 0.00308 \\
         & RGB & 259 & 198 & 52 & 0.00254 & 0.00215 & 0.00089 & 0.16000 \\
         & CHeB & 142 & 10 & 124 & 0.00607 & 0.00607 & 0.00212 & 0.38154 \\
        EAGB & RGB & 34 & 31 & 3 & 0.00102 & 0.00012 & 0.00005 & 0.00923 \\
         & CHeB & 35 & 24 & 11 & 0.00373 & 0.00054 & 0.00019 & 0.03385 \\

        \bottomrule
    \end{tabular}
\end{table*}

\begin{figure}
    \centering
    \includegraphics[width=\linewidth]{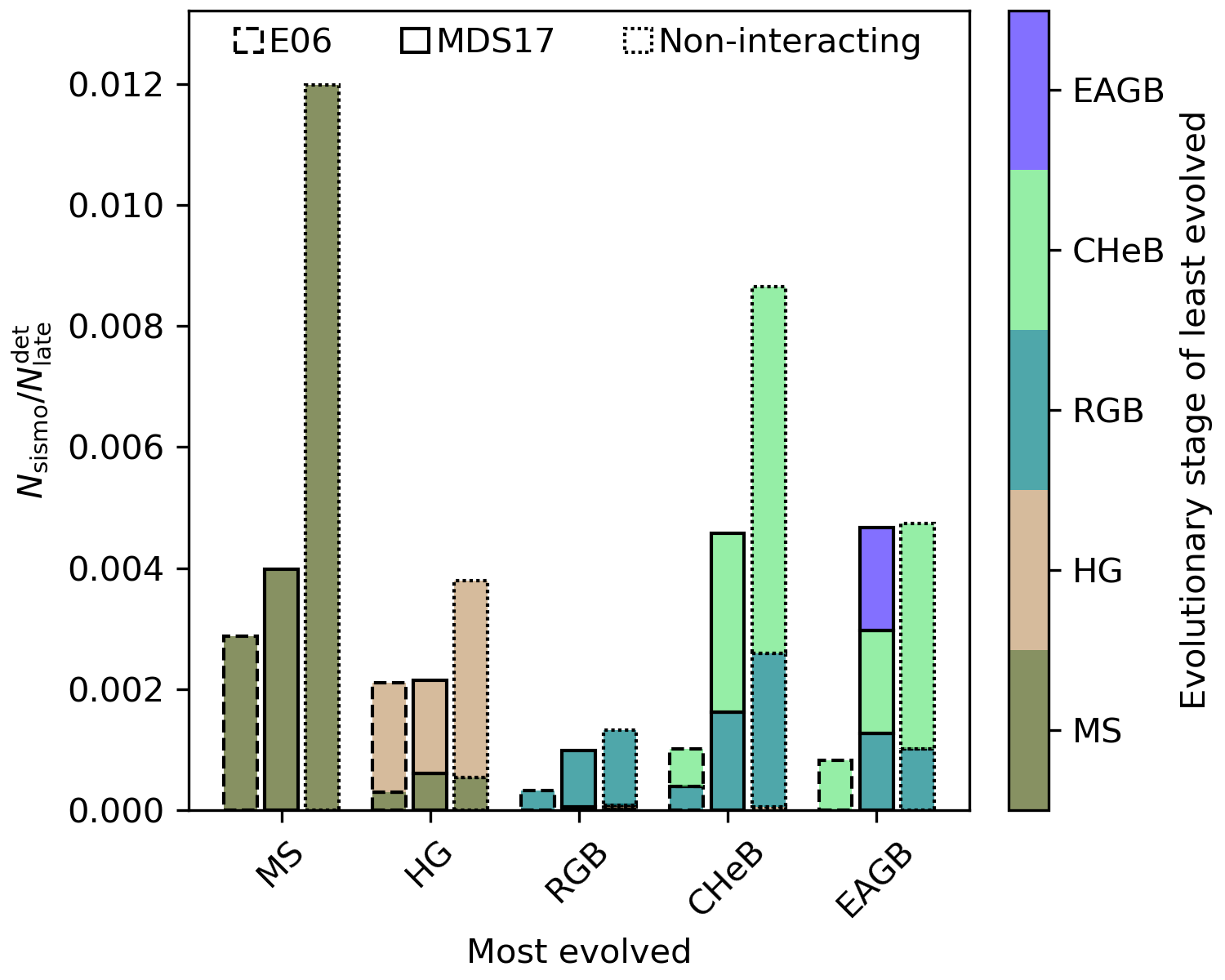}
    \caption{Comparison of the counts of different kinds of asteroseimic binaries in the three binary prescriptions we have used. The line styles indicate the prescription (E06 is dashed, MDS17 is continuous and non-interacting is dash-dotted), while the color stands for the evolutionary phase of the least evolved component of the binary. The evolutionary phase of the most evolved is indicated on the x-axis.}
    \label{fig:hist_detectable_bin_compare}
\end{figure}

It is worthwhile to explore the impact that different prescriptions for the initial parameters of binary systems and the inclusion of binary interactions have on the resulting counts of binary stars. In the following, we describe the differences between the MDS17, E06 and non-interacting binaries simulations for the asteroseismic binaries and the under-/over-massive stars.
Appendix~\ref{app:tables} provides tables of counts for all the kinds of binaries produced in each simulation.

\paragraph{Asteroseismic binaries.}
Figure~\ref{fig:hist_detectable_bin_compare} compares the fraction of asteroseismic binaries relative to the single stars with detectable oscillation in the evolutionary phase of the primary between the MDS17, E06 and non-interacting binaries simulations (Table~\ref{tab:counts_binaries} and Table~\ref{tab:counts_binaries_compare}).
A general trend appears: the E06 prescription leads to the lowest fraction of asteroseismic binaries, while the simulation with non-interacting binaries, as expected due to all binary systems surviving the evolution, produces the largest fraction.
The MDS17 simulation is similar to E06 one where the most evolved star is in the early stages of the evolution, and in later phases shows a much larger fraction, getting closer to the non-interacting case where the most evolved star is in the E-AGB phase.

The difference in the rates of asteroseismic binaries between the MDS17 and E06 simulations is mainly due to the different ranges of the initial period and eccentricity of the binaries.
While in the MDS17 prescription the period is limited to $0.2 < \log P (\si{\day}) < 8.0$ and the eccentricity has an upper limit that depends on the period, $0.0<e<e_\text{max}(P)$, the E06 prescription (Eq.~\ref{eq:egg_P}) has no sharp cut on period and eccentricity.

\paragraph{Under-/Over-massive stars}

\begin{table*}[t]
    \caption{Under-massive stars in the case of the E06 binary prescription. See also Table~\ref{tab:counts_under-massive}.}
    \label{tab:counts_under-massive_Egg}
    \centering
    \begin{tabular}{lrrrrrrrr}
        \toprule
        Phase &
        N &
        N$_\text{lone pri}$ &
        N$_\text{sec}$ &
        N$_\text{lone pri}^\text{det}$ &
        N$_\text{sec}^\text{det}$ &
        N$_\text{lone pri}^\text{det}$ / N$^\text{det}$ &
        N$_\text{sec}^\text{det}$ / N$^\text{det}$ &
        N$^\text{det}$ \\
        \midrule
        13 & 0 & 13 & 0 & 0 & 0.00000 & 0.00000 & 7671 \\
39 & 0 & 39 & 0 & 2 & 0.00000 & 0.00060 & 3323 \\
50 & 16 & 34 & 16 & 14 & 0.00074 & 0.00065 & 21568 \\
169 & 167 & 2 & 164 & 1 & 0.00919 & 0.00006 & 17843 \\
16 & 16 & 0 & 16 & 0 & 0.00658 & 0.00000 & 2433 \\
    \bottomrule
    \end{tabular}
\end{table*}
\begin{table*}[t]
    \caption{Over-massive stars in the case of the E06 binary prescription. See also Table~\ref{tab:counts_over-massive}.}
    \label{tab:counts_over-massive_Egg}
    \centering
    \begin{tabular}{lrrrrrrrrrrr}
        \toprule
        Phase &
        N &
        N$_\text{lone pri}$ &
        N$_\text{pri}$  &
        N$_\text{sec}$  &
        N$_\text{lone pri}^\text{det}$ &
        N$_\text{pri}^\text{det}$ &
        N$_\text{sec}^\text{det}$ &
        N$_\text{lone pri}^\text{det}$ / N$^\text{det}$ &
        N$_\text{pri}^\text{det}$ / N$^\text{det}$ &
        N$_\text{sec}^\text{det}$ / N$^\text{det}$ &
        N$^\text{det}$ \\
        \midrule
        MS & 4288 & 4194 & 83 & 11 & 189 & 0 & 0 & 0.02464 & 0.00000 & 0.00000 & 7671 \\
        HG & 540 & 534 & 6 & 0 & 69 & 0 & 0 & 0.02076 & 0.00000 & 0.00000 & 3323 \\
        RGB & 777 & 773 & 3 & 1 & 508 & 2 & 1 & 0.02355 & 0.00009 & 0.00005 & 21568 \\
        CHeB & 738 & 738 & 0 & 0 & 701 & 0 & 0 & 0.03929 & 0.00000 & 0.00000 & 17843 \\
        EAGB & 103 & 103 & 0 & 0 & 103 & 0 & 0 & 0.04233 & 0.00000 & 0.00000 & 2433 \\
        \bottomrule
    \end{tabular}
\end{table*}

\begin{figure}[t]
    \centering
    \includegraphics[width=\linewidth]{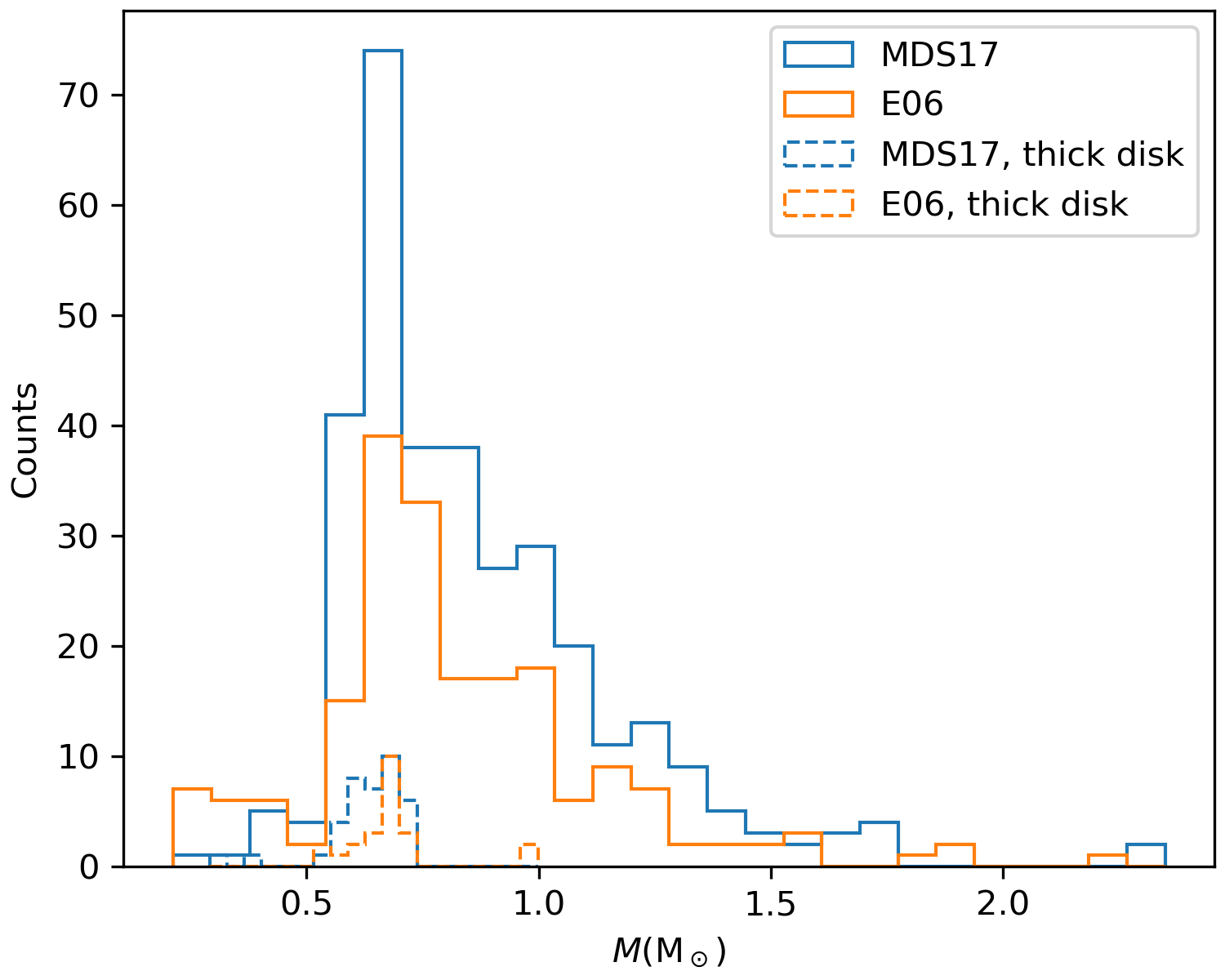}
    \caption{Distribution of masses of under-massive stars with detectable oscillations for the MDS17 and E06 simulations. Solid lines refer to the entire sample of each prescription, while dashed lines refer to stars in the thick disk of our Galaxy.}
    \label{fig:hist_under-massive_compare}
\end{figure}
\begin{figure}[t]
    \centering
    \includegraphics[width=\linewidth]{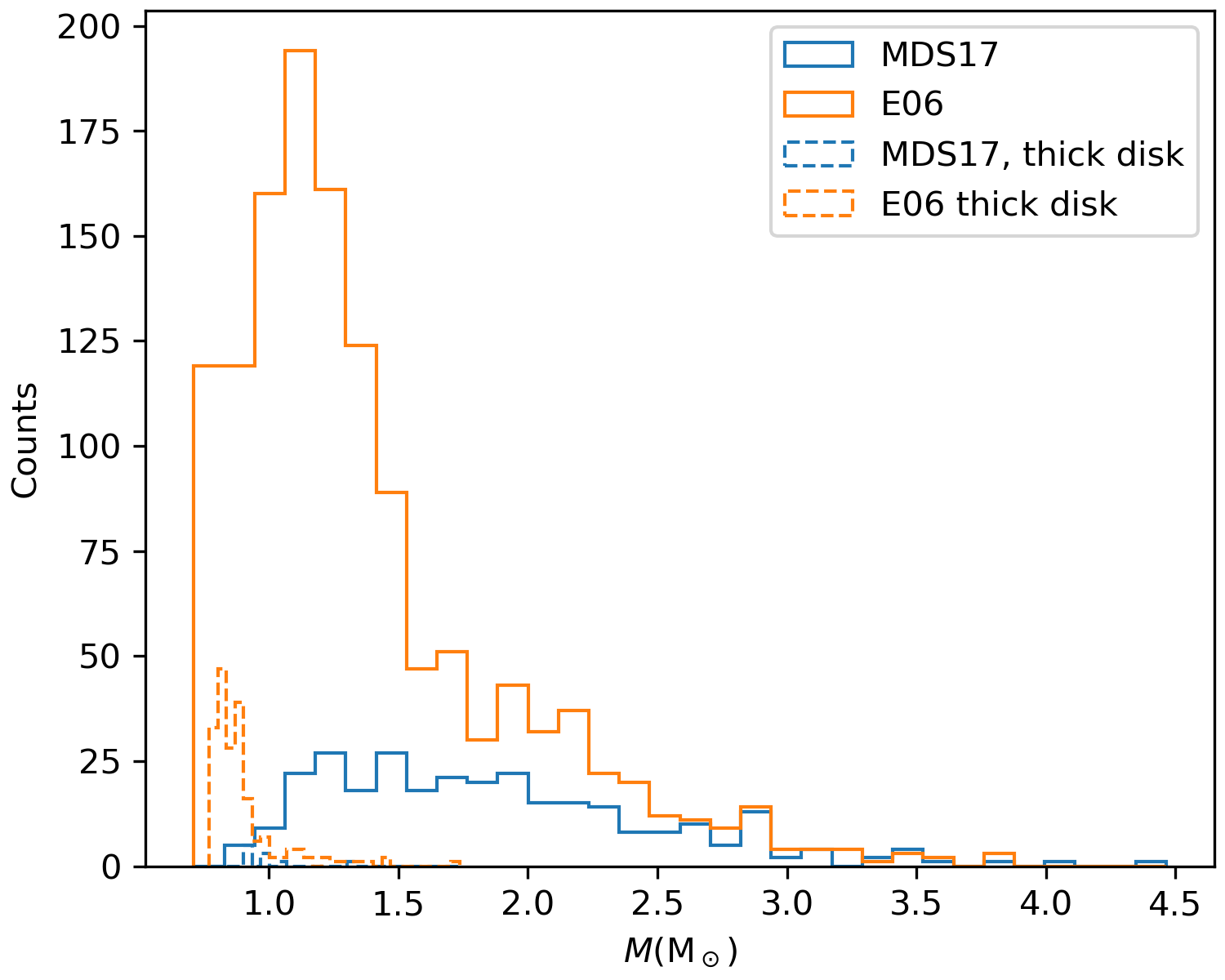}
    \caption{Distribution of masses of over-massive stars with detectable oscillations for the MDS17 and E06 simulations. Solid lines refer to the entire sample of each prescription, while dashed lines refer to stars in the thick disk of our Galaxy.}
    \label{fig:hist_over-massive_compare}
\end{figure}

As in the respective Sections for the MDS17 simulation (\ref{sec:results-under} and \ref{sec:results-over}), we select under-massive and over-massive stars from the E06 simulation (Table~\ref{tab:counts_under-massive_Egg} and Table~\ref{tab:counts_over-massive_Egg}) and compare their mass distributions with the MDS17 one.
For this discussion, we only focus on the evolved stars with detectable oscillations, and exclude all others.

Figure~\ref{fig:hist_under-massive_compare} compares the mass distribution of the under-massive stars for the MDS17 and E06 simulations.
These distributions have a similar shape and show a peak at a mass of $\SI{0.67}{\Msun}$, but the MDS17 prescription produces more under-massive stars.
If we restrict the sample to stars in the thick disk (dashed lines in Figure~\ref{fig:hist_under-massive_compare}), although the counts are very low, they show again a peak at the same location.

In Figure~\ref{fig:hist_over-massive_compare} we compare the mass distributions of over-massive stars.
The E06 simulation produces 9 times more stars with $M<\SI{1.5}{\Msun}$ and $\sim 2$ times more stars with $M>\SI{1.5}{\Msun}$ than the MDS17 one.
This is again due to the differences in the initial parameters of the binaries.
The E06 prescriptions generates more systems with short periods that interact and merge early in the evolution, leading to an increase in the number of over-massive stars especially in the low-mass ($M \lesssim \SI{1.5}{\Msun})$ regime.
However, these over-massive stars born on the MS will have a subsequent evolution similar to a star that was not subject to mass transfer events.

Limiting the comparison to the Galactic thick disk, the MDS17 simulation contains 27 over-massive stars, of which 1 RGB, 8 CHeB and 1 EAGB stars have detectable oscillations, while the E06 one produces 410 over-massive stars up to the EAGB, of which 80 RGB and 109 CHeB have detectable oscillations.
In the region of the RGB below the RC ($\log g=2.4$ and $\log g=3.1$), the E06 simulation produces $30$ over-massive RGB stars with detectable oscillations, whilst MDS17 produces none.
For every 1000 RGB stars with detectable oscillations in the thick disk, with the MDS17 prescription we expect $0.3$ over-massive RGB with detectable oscillations, while with the E06 one we expect $20$.
In the case of CHeB stars, for every 1000 with detectable oscillations, the MDS17 prescription produces $2.7$ over-massive CHeB with detectable oscillations, and the E06 one $38$.

In the MDS17 simulation (Table~\ref{tab:counts_over-massive}), we can see that the fraction of over-massive giants with detectable oscillations relative to the total number of red giants with detectable oscillations is $0.7\%$, while for the E06 one (Table~\ref{tab:counts_over-massive_Egg}) it is about $3\%$.
Previous studies, however, found a larger fraction of over-massive stars.
\cite{brogaard16} report that in the open cluster NGC~6819 there is a fraction of over-massive giant stars of around $10\%$ or larger.
\cite{miglio21}, using \Kepler{} and APOGEE DR14 data, observed that the fraction of over-massive stars on the RGB between $\log g=2.4$ and $\log g=3.1$, is $\sim5\%$, while for CHeB stars it is $\sim18 \%$.
More recently, \cite{grisoni24} analyzed K2 and APOGEE data for about 6000 stars and found that YAR stars counts are 7-10\% of the total counts for the high-$\alpha$ stars.

\cite{izzard18} has gone into detail about predicting over-massive giant stars in the thick disk, defined in their simulations as any star with $M>\SI{1.3}{\Msun}$, and testing the effects of different assumptions in the models.
They found that these stars constitute between 1\% and 3\% of all the giants, but changing the distribution of initial separations makes them 10\% of the total.

\section{Finding binaries and binary products}
\label{sec:finding}

Given the effect that different prescriptions have on the resulting synthetic populations of binaries, it is important to collect observational evidence, that is, to identify and characterize as many binaries as possible.
There are multiple techniques to perform this task, each covering a slightly different region of the parameter space \citep[see for instance][]{moedistefano2017}.

In particular, with the advent of the Gaia mission, the search for binary systems has received a substantial push forward, as Gaia's astrometry, photometry and spectroscopy have been shown to be extremely powerful tools \citep{arenou23_gaiaDR3_multiplicity}.
For instance, the motion of the components of the binary system around the center of mass leads to a motion of the photometric center of the system that cannot be fitted with a single star astrometric model and sometimes results in a high \texttt{ruwe} (Renormalised Unit Weight Error) value.

\cite{elbadry21} used Gaia eDR3 to make a catalog of wide binaries within \SI{1}{\kilo\pc} of the Sun; \cite{penoyre22_ii} looked for binaries in the Gaia Catalog of Nearby Stars \citep{GCNSs21} using a renormalized version of \texttt{ruwe}; more recently, \cite{castroginard24} investigated the probability of detecting unresolved binary systems using a \texttt{ruwe} threshold that depends on the location on the sky.
However, a limitation of this kind of analysis is that as the distance increases, the signal that one can expect from the \texttt{ruwe} decreases as well.
In fact, \cite{castroginard24} show that already at \SI{1}{\kilo\pc} the probability of detecting a binary system reaches 50\%, and it is also a strong function of the period.

Asteroseismology can fill in with the characterization of binaries, as it can determine the intrinsic properties of stars under the condition that a long enough photometric time-series is available, and is largely independent from the orbital period of the binary.
Asteroseismic binaries, where both components have detectable oscillations, are the perfect target to study, although disentangling the spectra can be challenging.

Even binaries where only one component has detectable oscillations may be identified.
Indeed, asteroseismology can provide asteroseismic radii and, if photospheric constraints on effective temperature (and possibly metallicity) can be obtained from spectroscopy, an expected luminosity can be computed.
Comparing this value with observed magnitudes then yields a distance of the source \citep[see, e.g.][]{rodrigues14, khan19, khan23_i}.
For binary stars where the Gaia parallax is available, and the motion of the photocentre does not significantly affect its estimation, a strong mismatch with an asteroseismically derived distance might suggest that the source is, in fact, a binary star.

\begin{figure}
    \centering
    \includegraphics[width=\linewidth]{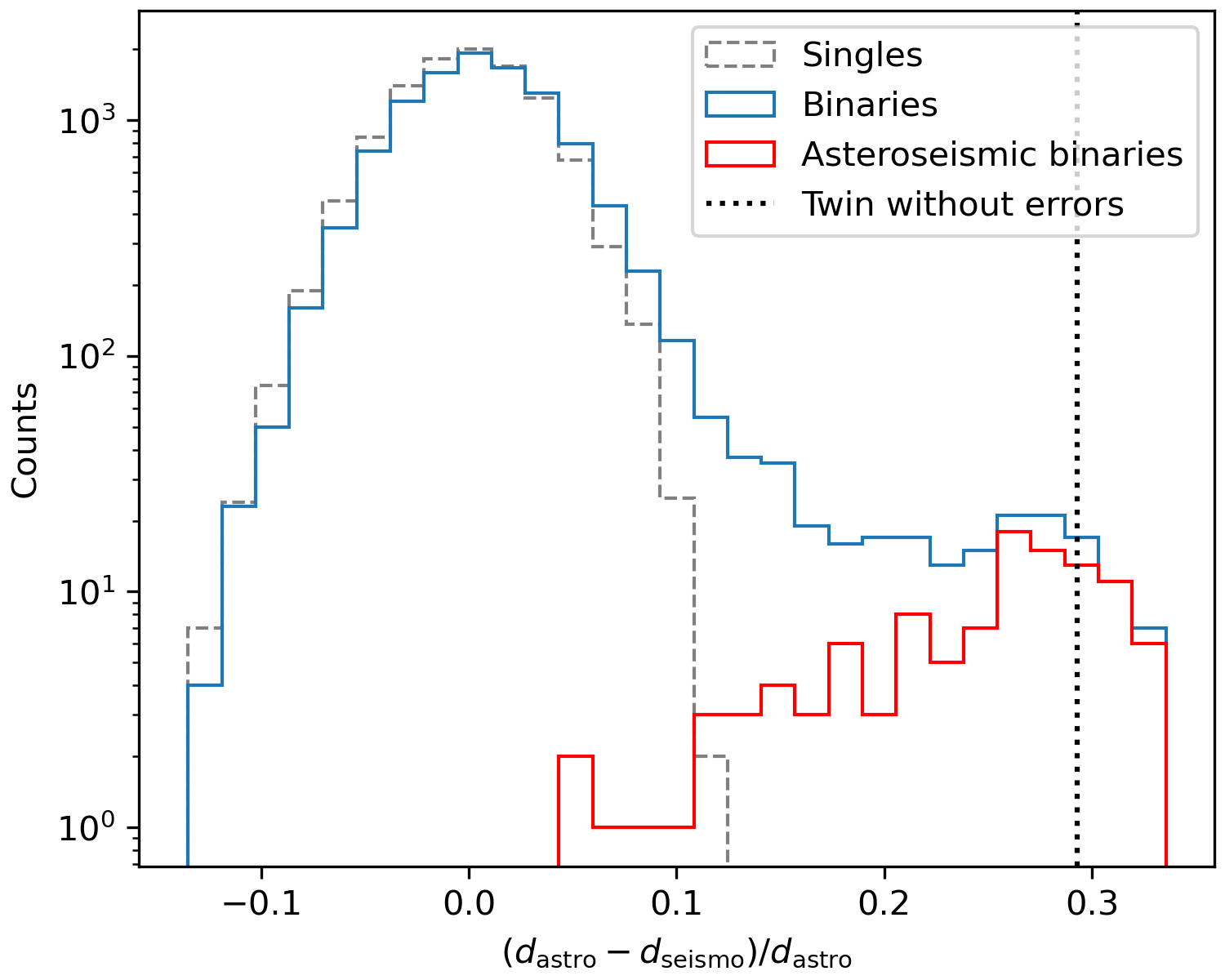}
    \caption{Toy model showing the effect on distance determination of unresolved binaries. The gray dashed line shows the reference case, made of only single stars. The blue line shows the effect on distance determination for binaries and the red line highlights the effect on asteroseismic binaries. The black dotted line indicates the expected relative distance difference in the case of a twin, i.e. a binary where $L_1=L_2$, ignoring the effect of measurement errors.}
    \label{fig:distance_comparison}
\end{figure}

To show this effect we have built a simple toy model. We take the distances of the stars and binaries in the simulation and add to them a normally distributed error with $\sigma=2\%$ to represent a distance estimate obtained from astrometry, $d_\text{astro}$, assuming it is large enough to ignore the motion of the photocentre.
Then, we compute the intrinsic luminosity of the brightest component of each binary using the radius, which in a real application would come from asteroseismology, $R_\text{seismo}$, and the effective temperature $T_\text{eff}$:
\begin{equation}
    L(\si{\Lsun}) = \left( \frac{ \hat{R}_\text{seismo} }{ \si{\Rsun} } \right)^2 \left( \frac{ \hat{T}_\text{eff}}{ \si{\Tsun} } \right)^4 \quad,
\end{equation}
where all quantities marked with a hat are taken from the simulation and perturbed with a 2\% gaussian error to represent uncertainties in the measurements.
Finally, we compute the apparent total luminosity of the unresolved binary $l_\text{tot}$ and compare it to the intrinsic one to derive an asteroseismically informed distance:
\begin{equation}
    d_\text{seismo}^{2} = \frac{L}{l_\text{tot}} \quad .
\end{equation}

We show in Fig.~\ref{fig:distance_comparison} the results of this very simple model as a histogram of relative differences between astrometric and asteroseismic distance.
If all stars were single (gray line), the distribution of the differences would be symmetric around zero.
Instead, a population of only binaries (blue line) would exhibit a tail at large relative differences, meaning that they appear to be closer than the astrometric distance suggests.
Asteroseismic binaries (red line), having a mass ratio close to one and a similar luminosity ratio, would represent most of the systems with largely underestimated distances.
Without errors on any of the quantities involved in the toy model and considering a twin binary system where $L_1=L_2$, we would expect $\left(d_\text{astro}-d_\text{seismo}\right)/d_\text{astro} = 1 - \sqrt{2}/2 \approx 0.29$ (black dotted line in the Figure).
Some binary systems show a larger $\left(d_\text{astro}-d_\text{seismo}\right)/d_\text{astro}$ owing to the measurement errors.

Nevertheless, we note that contamination can also lead to large $\left(d_\text{astro}-d_\text{seismo}\right)/d_\text{astro}$, and can be particularly important for large pixels sizes as in the case of the TESS mission \citep[Transiting Exoplanet Survey Satellite, ][]{ricker14_tess}, and the upcoming mission PLATO \cite[PLAnetary Transits and Oscillations of stars, ][]{rauer24_plato}.

Missions like \Kepler{}, K2 and TESS provide a vast collection of light curves for identifying binary stars, and future missions with asteroseismic capabilities could push even further the boundaries.

\section{Conclusions}
\label{sec:conclusion}

In this work, we present and discuss three simulations of the content of \Kepler{}'s field of view using different binary prescriptions: interacting binaries with the \cite{eggleton06} and \cite{moedistefano2017} parameter distributions, and non-interacting binaries.
We use the TRILEGAL population synthesis code and its BinaPSE module to account for the effects of binary evolution and binary interactions in the simulated stellar populations, and analyze the results taking into account the detectability of solar-like oscillations of the simulated systems.
Although the simulations we have run use $\fbin=0.3$ (Section~\ref{sec:sim-setup}), results for a different value of $\fbin$ can be roughly estimated by linearly rescaling the counts presented in this work.

We summarize our findings in the following.
\begin{itemize}
    \item The simulation adopting the MDS17 distribution of initial parameters produces a small but significant number of asteroseismic binaries, e.g. about $1$ double RGB for every 1000 RGB stars with detectable oscillations and $2$ composed of a CHeB and an RGB every 1000 CHeB stars with detectable oscillations (Table~\ref{tab:counts_binaries}). A large fraction of the asteroseismic binaries are expected to consist of double CHeB binaries ($35\%$), followed by mixed CHeB and RGB binaries ($19\%$) and double RGB binaries ($13\%$).
    \item Asteroseismic binaries are expected to not have exchanged mass during their evolution and therefore to retain their initial mass ratio, which should be close to one except for cases where one component has lost mass to single stellar evolution. In the MDS17 simulation (Figure~\ref{fig:mass_ratio}), $95\%$ of the asteroseismic binaries have an initial mass ratio larger than $0.96$. The initial semi-major axis of asteroseismic binaries should be conserved as well, although circularization can take place (Figure~\ref{fig:semimajoraxis}).
    \item The occurrence of asteroseismic double CHeB binaries is strongly influenced by interaction (Fig.~\ref{fig:hist_detectable_bin_compare}). With the MDS17 prescription, we can find $3$ for every 1000 CHeB with detectable oscillations, while with the E06 one we find $0.6$ and in the non-interacting case about $6$ (Table~\ref{tab:counts_binaries_compare}).
    \item Double RC binaries ($M_1+M_2\lesssim\SI{4}{\Msun}$ in Fig.\ref{fig:double_rc_sma_mtot}) are not expected to survive binary evolution if their orbital separation is close to or smaller than their Roche lobe at the RGB tip (Figure~\ref{fig:double_rc_sma_mtot}), that is for semi-major axis shorter than about \SI{500}{\Rsun} or for initial orbital periods shorter than about \SI{1000}{\day}. Larger mass binaries are not subject to a comparable expansion on the RGB and can be found at much smaller orbital separations.
    \item About $8$ under-massive and $7$ over-massive red giant stars with detectable oscillations are expected to be found for every 1000 red giant stars with detectable oscillations in the MDS17 simulation. Both kinds of stars would have apparent (derived) ages different from what would be expected for their metallicity.
    \item In the E06 simulation, for every 1000 red giant stars with detectable oscillations, $5$ are under-massive (Table~\ref{tab:counts_under-massive_Egg}), and $31$ are over-massive (Table~\ref{tab:counts_over-massive_Egg}). However, this count also includes numerous low-mass systems that interacted very early in their evolution and in later phases appear as regular stars, that is they ultimately have a mass compatible with their age.
\end{itemize}

In conclusion, varying assumptions about the initial binary parameter distributions give rise to subtle yet significant differences in stellar populations, which directly affect our ability to accurately trace the history of our galaxy.
Under-massive stars, whether on the turnoff or in the giant branches, can lead to significant age overestimation if not correctly identified, while over-massive stars may appear much younger than their true age.
Both scenarios, if not properly considered, can introduce additional scatter in the age-metallicity relation (AMR) that is measured \citep{izzard18, miglio21}.

These objects are more easily detected in stellar clusters \citep[e.g.][]{leiner16, handberg17, brogaard21_ngc6791, matteuzzi24}, where all stars have the same age and initial chemical composition.
In the field, however, under- and over-massive stars are much more difficult to identify.

To reduce uncertainties in the models, such as those related to mass transfer, and to better constrain assumptions on binary evolution such as the initial distribution of semi-major axis and the initial binary fraction, a systematic comparison between simulated and observed binaries and products of their evolution is necessary.
The simulations we have produced serve as a foundation for such an investigation in \Kepler{} data, providing insight into the types of systems we can expect to detect.
Furthermore, this study will assist in the preparation for the PLATO mission, and will aid in the interpretation of the binary stellar populations found with its observations.

\begin{acknowledgements}
A.M. and J.S.T. acknowledges financial support from Bologna University, ``MUR FARE Grant Duets CUP J33C21000410001''.
A.Miglio, KB, MM, and WEvR acknowledge support from the ERC Consolidator Grant funding scheme (project ASTEROCHRONOMETRY, G.A. n. 772293 \url{http://www.asterochronometry.eu}).
D.B.~acknowledges funding support by the Italian Ministerial Grant PRIN 2022, ``Radiative opacities for astrophysical applications'', no.~2022NEXMP8, CUP C53D23001220006.
Some of the results in this paper have been derived using the healpy and HEALPix packages.
The simulations described in this work are available upon reasonable request to the authors.

\end{acknowledgements}

\bibliographystyle{aa}
\bibliography{references}

\label{LastPage}

\begin{appendix}

\onecolumn

\section{Full tables of binary star counts}\label{app:tables}

Tables~\ref{tab:full_binary_counts_MDS}, \ref{tab:full_binary_counts_E} and \ref{tab:full_binary_counts_non_interacting} present the counts of all binaries produced by the MDS17, E06 and non-interacting binaries simulations, respectively.

\begin{table}[h!]
    \caption{Binary systems generated by the MDS17 simulation. Similar to \ref{tab:counts_binaries}, but extended also to combinations of primary and secondary phases that do not result in any asteroseismic binaries. Every binary in the table has an unresolved magnitude brighter than \SI{16}{\mag} in \Kepler{}'s band.}
    \label{tab:full_binary_counts_MDS}
    \centering
        \begin{tabular}{llrrrrrrrrr}
            \toprule
            phase$_\text{late}$ & phase$_\text{early}$ & N & N$_\text{one}$ & N$_\text{seismo}$ & N$_\text{seismo}$ / N$_\text{late}^\text{det}$ & N$_\text{seismo}$  / N$_\text{early}^\text{det}$ & N$_\text{seismo}$ / $\sum$ N$_\text{late}^\text{det}$ & N$_\text{seismo}$/$\sum$N$_\text{seismo}$ \\
            \midrule
            MS & MS & 59228 & 2070 & 30 & 0.19868 & 0.00399 & 0.00399 & 0.00057 \\
            HG & MS & 7620 & 1023 & 2 & 0.01325 & 0.00061 & 0.00027 & 0.00004 \\
             & HG & 521 & 9 & 5 & 0.03311 & 0.00153 & 0.00153 & 0.00010 \\
            RGB & MS & 8738 & 5589 & 1 & 0.00662 & 0.00005 & 0.00013 & 0.00002 \\
             & HG & 484 & 194 & 0 & 0.00000 & 0.00000 & 0.00000 & 0.00000 \\
             & RGB & 257 & 133 & 20 & 0.13245 & 0.00094 & 0.00094 & 0.00038 \\
            CHeB & MS & 4975 & 4581 & 0 & 0.00000 & 0.00000 & 0.00000 & 0.00000 \\
             & HG & 123 & 43 & 0 & 0.00000 & 0.00000 & 0.00000 & 0.00000 \\
             & RGB & 145 & 69 & 29 & 0.19205 & 0.00162 & 0.00137 & 0.00055 \\
             & CHeB & 65 & 2 & 53 & 0.35099 & 0.00296 & 0.00296 & 0.00101 \\
            EAGB & MS & 518 & 510 & 0 & 0.00000 & 0.00000 & 0.00000 & 0.00000 \\
             & HG & 14 & 13 & 0 & 0.00000 & 0.00000 & 0.00000 & 0.00000 \\
             & RGB & 12 & 8 & 3 & 0.01987 & 0.00127 & 0.00014 & 0.00006 \\
             & CHeB & 6 & 2 & 4 & 0.02649 & 0.00170 & 0.00022 & 0.00008 \\
             & EAGB & 4 & 0 & 4 & 0.02649 & 0.00170 & 0.00170 & 0.00008 \\
            TP-AGB & MS & 16 & 16 & 0 & 0.00000 & 0.00000 & 0.00000 & 0.00000 \\
            Post-AGB & MS & 20 & 17 & 0 & 0.00000 & 0.00000 & 0.00000 & 0.00000 \\
             & RGB & 1 & 1 & 0 & 0.00000 & 0.00000 & 0.00000 & 0.00000 \\
            CO-WD & MS & 3570 & 0 & 0 & 0.00000 & 0.00000 & 0.00000 & 0.00000 \\
             & HG & 1026 & 1 & 0 & 0.00000 & 0.00000 & 0.00000 & 0.00000 \\
             & RGB & 1893 & 13 & 0 & 0.00000 & 0.00000 & 0.00000 & 0.00000 \\
             & CHeB & 1271 & 30 & 0 & 0.00000 & 0.00000 & 0.00000 & 0.00000 \\
             & EAGB & 120 & 4 & 0 & 0.00000 & 0.00000 & 0.00000 & 0.00000 \\
             & TP-AGB & 2 & 0 & 0 & 0.00000 & 0.00000 & 0.00000 & 0.00000 \\
             & Post-AGB & 12 & 0 & 0 & 0.00000 & 0.00000 & 0.00000 & 0.00000 \\
            He-WD & MS & 5647 & 0 & 0 & 0.00000 & 0.00000 & 0.00000 & 0.00000 \\
             & HG & 972 & 2 & 0 & 0.00000 & 0.00000 & 0.00000 & 0.00000 \\
             & RGB & 973 & 9 & 0 & 0.00000 & 0.00000 & 0.00000 & 0.00000 \\
             & CHeB & 405 & 3 & 0 & 0.00000 & 0.00000 & 0.00000 & 0.00000 \\
             & EAGB & 44 & 0 & 0 & 0.00000 & 0.00000 & 0.00000 & 0.00000 \\
            NS & MS & 463 & 14 & 0 & 0.00000 & 0.00000 & 0.00000 & 0.00000 \\
             & HG & 65 & 9 & 0 & 0.00000 & 0.00000 & 0.00000 & 0.00000 \\
             & RGB & 107 & 63 & 0 & 0.00000 & 0.00000 & 0.00000 & 0.00000 \\
             & CHeB & 81 & 45 & 0 & 0.00000 & 0.00000 & 0.00000 & 0.00000 \\
             & EAGB & 7 & 6 & 0 & 0.00000 & 0.00000 & 0.00000 & 0.00000 \\
             & Post-AGB & 2 & 2 & 0 & 0.00000 & 0.00000 & 0.00000 & 0.00000 \\
            BH & MS & 5 & 0 & 0 & 0.00000 & 0.00000 & 0.00000 & 0.00000 \\
            \bottomrule
        \end{tabular}
\end{table}

\begin{table}
    \caption{Similar to \ref{tab:full_binary_counts_MDS}, but for the E06 simulation.}
    \label{tab:full_binary_counts_E}
    \centering
        \begin{tabular}{llrrrrrrrrr}
            \toprule
            phase$_\text{late}$ & phase$_\text{early}$ & N & N$_\text{one}$ & N$_\text{seismo}$ & N$_\text{seismo}$ / N$_\text{late}^\text{det}$ & N$_\text{seismo}$  / N$_\text{early}^\text{det}$ & N$_\text{seismo}$ / $\sum$ N$_\text{late}^\text{det}$ & N$_\text{seismo}$/$\sum$N$_\text{seismo}$ \\
            \midrule
            MS & MS & 55691 & 1908 & 22 & 0.39286 & 0.00287 & 0.00287 & 0.00041 \\
            HG & MS & 7275 & 913 & 1 & 0.01786 & 0.00030 & 0.00013 & 0.00002 \\
             & HG & 253 & 6 & 6 & 0.10714 & 0.00181 & 0.00181 & 0.00011 \\
            RGB & MS & 8792 & 5671 & 0 & 0.00000 & 0.00000 & 0.00000 & 0.00000 \\
             & HG & 277 & 123 & 0 & 0.00000 & 0.00000 & 0.00000 & 0.00000 \\
             & RGB & 87 & 38 & 7 & 0.12500 & 0.00032 & 0.00032 & 0.00013 \\
            CHeB & MS & 4996 & 4610 & 0 & 0.00000 & 0.00000 & 0.00000 & 0.00000 \\
             & HG & 74 & 35 & 0 & 0.00000 & 0.00000 & 0.00000 & 0.00000 \\
             & RGB & 31 & 15 & 7 & 0.12500 & 0.00039 & 0.00032 & 0.00013 \\
             & CHeB & 12 & 0 & 11 & 0.19643 & 0.00062 & 0.00062 & 0.00021 \\
            EAGB & MS & 551 & 548 & 0 & 0.00000 & 0.00000 & 0.00000 & 0.00000 \\
             & HG & 6 & 6 & 0 & 0.00000 & 0.00000 & 0.00000 & 0.00000 \\
             & RGB & 3 & 3 & 0 & 0.00000 & 0.00000 & 0.00000 & 0.00000 \\
             & CHeB & 3 & 1 & 2 & 0.03571 & 0.00082 & 0.00011 & 0.00004 \\
            TP-AGB & MS & 15 & 15 & 0 & 0.00000 & 0.00000 & 0.00000 & 0.00000 \\
            Post-AGB & MS & 14 & 13 & 0 & 0.00000 & 0.00000 & 0.00000 & 0.00000 \\
            CO-WD & MS & 3448 & 0 & 0 & 0.00000 & 0.00000 & 0.00000 & 0.00000 \\
             & HG & 1043 & 0 & 0 & 0.00000 & 0.00000 & 0.00000 & 0.00000 \\
             & RGB & 1706 & 5 & 0 & 0.00000 & 0.00000 & 0.00000 & 0.00000 \\
             & CHeB & 1141 & 13 & 0 & 0.00000 & 0.00000 & 0.00000 & 0.00000 \\
             & EAGB & 108 & 2 & 0 & 0.00000 & 0.00000 & 0.00000 & 0.00000 \\
             & TP-AGB & 5 & 1 & 0 & 0.00000 & 0.00000 & 0.00000 & 0.00000 \\
             & Post-AGB & 1 & 0 & 0 & 0.00000 & 0.00000 & 0.00000 & 0.00000 \\
             & CO-WD & 1 & 0 & 0 & 0.00000 & 0.00000 & 0.00000 & 0.00000 \\
            He-WD & MS & 6097 & 1 & 0 & 0.00000 & 0.00000 & 0.00000 & 0.00000 \\
             & HG & 1092 & 1 & 0 & 0.00000 & 0.00000 & 0.00000 & 0.00000 \\
             & RGB & 1073 & 5 & 0 & 0.00000 & 0.00000 & 0.00000 & 0.00000 \\
             & CHeB & 455 & 2 & 0 & 0.00000 & 0.00000 & 0.00000 & 0.00000 \\
             & EAGB & 33 & 1 & 0 & 0.00000 & 0.00000 & 0.00000 & 0.00000 \\
             & TP-AGB & 2 & 0 & 0 & 0.00000 & 0.00000 & 0.00000 & 0.00000 \\
            NS & MS & 129 & 6 & 0 & 0.00000 & 0.00000 & 0.00000 & 0.00000 \\
             & HG & 13 & 3 & 0 & 0.00000 & 0.00000 & 0.00000 & 0.00000 \\
             & RGB & 17 & 9 & 0 & 0.00000 & 0.00000 & 0.00000 & 0.00000 \\
             & CHeB & 24 & 10 & 0 & 0.00000 & 0.00000 & 0.00000 & 0.00000 \\
             & EAGB & 1 & 0 & 0 & 0.00000 & 0.00000 & 0.00000 & 0.00000 \\
            BH & RGB & 1 & 1 & 0 & 0.00000 & 0.00000 & 0.00000 & 0.00000 \\
            \bottomrule
        \end{tabular}
\end{table}

\begin{table}
    \caption{Similar to \ref{tab:full_binary_counts_MDS}, but for the simulation with non-interacting binaries.}
    \label{tab:full_binary_counts_non_interacting}
    \centering
    \begin{tabular}{llrrrrrrrrr}
        \toprule
        phase$_\text{late}$ & phase$_\text{early}$ & N & N$_\text{one}$ & N$_\text{seismo}$ & N$_\text{seismo}$ / N$_\text{late}^\text{det}$ & N$_\text{seismo}$  / N$_\text{early}^\text{det}$ & N$_\text{seismo}$ / $\sum$ N$_\text{late}^\text{det}$ & N$_\text{seismo}$/$\sum$N$_\text{seismo}$ \\
        \midrule
        MS & MS & 62384 & 1150 & 86 & 0.26462 & 0.01199 & 0.01199 & 0.00147 \\
        HG & MS & 6899 & 599 & 2 & 0.00615 & 0.00054 & 0.00028 & 0.00003 \\
            & HG & 684 & 23 & 12 & 0.03692 & 0.00325 & 0.00325 & 0.00021 \\
        RGB & MS & 7953 & 4947 & 1 & 0.00308 & 0.00004 & 0.00014 & 0.00002 \\
            & HG & 998 & 523 & 1 & 0.00308 & 0.00004 & 0.00027 & 0.00002 \\
            & RGB & 351 & 202 & 30 & 0.09231 & 0.00124 & 0.00124 & 0.00051 \\
        CHeB & MS & 4267 & 4102 & 0 & 0.00000 & 0.00000 & 0.00000 & 0.00000 \\
            & HG & 385 & 376 & 1 & 0.00308 & 0.00005 & 0.00027 & 0.00002 \\
            & RGB & 259 & 198 & 52 & 0.16000 & 0.00254 & 0.00215 & 0.00089 \\
            & CHeB & 142 & 10 & 124 & 0.38154 & 0.00607 & 0.00607 & 0.00212 \\
        EAGB & MS & 597 & 597 & 0 & 0.00000 & 0.00000 & 0.00000 & 0.00000 \\
            & HG & 57 & 56 & 0 & 0.00000 & 0.00000 & 0.00000 & 0.00000 \\
            & RGB & 34 & 31 & 3 & 0.00923 & 0.00102 & 0.00012 & 0.00005 \\
            & CHeB & 35 & 24 & 11 & 0.03385 & 0.00373 & 0.00054 & 0.00019 \\
        Post-AGB & MS & 11239 & 549 & 0 & 0.00000 & 0.00000 & 0.00000 & 0.00000 \\
            & HG & 4059 & 621 & 0 & 0.00000 & 0.00000 & 0.00000 & 0.00000 \\
            & RGB & 6793 & 4579 & 0 & 0.00000 & 0.00000 & 0.00000 & 0.00000 \\
            & CHeB & 3934 & 3814 & 0 & 0.00000 & 0.00000 & 0.00000 & 0.00000 \\
            & EAGB & 635 & 632 & 0 & 0.00000 & 0.00000 & 0.00000 & 0.00000 \\
            & Post-AGB & 1010 & 0 & 0 & 0.00000 & 0.00000 & 0.00000 & 0.00000 \\
        \bottomrule
    \end{tabular}
\end{table}

\end{appendix}

\end{document}